\documentclass[floatfix,aps,tightenlines,prd,nofootinbib,superscriptaddress,twocolumn]{revtex4-1}

\usepackage{amsmath}
\usepackage{color}
\usepackage{graphicx}

\newcommand{\K}[3]{\ensuremath{ \mathrm{K}^{#1}_{\cdot \, #2 #3}}}
\newcommand{\T}[3]{\ensuremath{ \mathrm{T}^{#1}_{\cdot \, #2 #3}}}

\newcommand{\Ti}[3]{\ensuremath{\mathrm{T}^{\cdot \, #1}_{#2\, \cdot #3}}}
\newcommand{\Chris}[3]{\ensuremath{\Gamma^{#1}_{\cdot \, #2 #3}}}
\newcommand{\Christilde}[3]{\ensuremath{ \tilde{\Gamma}^{#1}_{\cdot \, #2 #3}}}
\newcommand{\Chrishat}[3]{\ensuremath{ \hat{\Gamma}^{#1}_{\cdot \, #2 #3}}}
\newcommand{\Chrisdash}[3]{\ensuremath{\bar{\Gamma}^{#1}_{\cdot \, #2 #3}}}
\newcommand{\Chrisbar}[3]{\ensuremath{ \bar{\Gamma}^{#1}_{\cdot \, #2 #3}}}
\newcommand{\A}{\ensuremath{\mathrm{A}}}
\newcommand{\J}{\ensuremath{\mathrm{J}}}
\newcommand{\F}{\ensuremath{\mathrm{F}}}

\newcommand{\h}{\ensuremath{\mathrm{H}}}
\newcommand{\B}{\ensuremath{\mathrm{B}}}
\newcommand{\G}{\ensuremath{\mathrm{G}}}
\newcommand{\g}{\ensuremath{\mathrm{g}}}

\begin{document}
\title{A metric theory of gravity with torsion in extra-dimension}
\author{Karthik H. Shankar}
\affiliation{Center for Memory and Brain, Boston University}
\author{Anand Balaraman}
\affiliation{Department of Physics, Georgia Southern University}
 \author{ Kameshwar C. Wali}
\affiliation{Department of Physics, Syracuse University}

\begin{abstract}
 We consider a theory of gravity with a hidden extra-dimension and metric-dependent torsion.  A set of physically motivated constraints are imposed on the geometry so that the torsion stays confined to the extra-dimension and the extra-dimension stays hidden at the level of four dimensional geodesic motion.  At the kinematic level, the theory maps on to General Relativity, but the dynamical field equations that follow from the action principle deviate markedly from the standard Einstein equations. We study static spherically symmetric vacuum solutions and homogeneous-isotropic cosmological solutions that emerge from the field equations. In both cases, we find solutions of significant physical interest. Most notably, we find positive mass solutions with naked singularity that match the well known Schwarzschild solution at large distances but lack an event horizon. In the cosmological context, we find oscillatory scenario in contrast to the inevitable singular big bang of the standard cosmology.
\end{abstract}

\maketitle{}

\section{Introduction}

Einstein viewed the space-time as a pseudo-Riemannian differentiable manifold in order to  generalize the special-relativistic flat space-time to include gravity. This was primarily motivated by the fact that the local flatness of the manifold structure naturally implemented his principle of equivalence. The generalization came along with the revolutionary idea that the trajectory of any freely moving test body is  simply a geodesic in the curved manifold, and that gravity  is not a Newtonian instantaneous action force, but an effect of the curvature of the space-time manifold. 

The basic constituents of the manifold structure are the \emph{metric}, that defines the distance between any two points of the manifold, and the \emph{connection}, that defines the covariant derivative and the curvature of the manifold.  Any theory of gravity should couple the dynamics of these quantities to the dynamics of the matter moving in the space-time manifold.  Among the existing theories, the ensuing field equations of general relativity (GR) are perhaps the simplest. Since in GR, \emph{torsion}, the antisymmetric combination of connection coefficients, is identically zero, and since GR  has withstood numerous precise experimental tests \cite{Will, Dicke}, introduction of torsion has seemed superfluous except in the presence of matter with intrinsic spin as in Einstein-Cartan formulations \cite{Hehl, Watanabe, Shapiro, Deser, Poplawski}. 

However, for two major reasons, alternate theories of gravity that reduce to GR in the weak field limit are seriously pursued. The first reason is that  GR leads to inevitable singularities - black holes (death of a massive star) and big bang (birth of the universe). Though it is conventionally assumed that quantization would eliminate these singularities, GR is not readily amenable to quantization. The second reason is that the standard model of cosmology based on GR requires most of the universe to be composed of unknown dark energy in order to account for various cosmological observations \cite{Trodden}. 
A common strategy to construct modified theories of gravity is to make the Lagrangian density  a nontrivial function of the Ricci scalar   \cite{SotiFara, Olmo} and use the action principle to derive the modified field equations.  Another common strategy  is to introduce extra-dimensions while constraining the physical particles to a (3+1) dimensional hyper-surface as in the brane-world theories \cite{Maartens, DGP, RandSund}. 
In this paper we explore a different approach by introducing metric-dependent torsion in Kaluza-Klein type theories  \cite{Wesson, Kalinowski} with one extra dimension.

In our approach, we consider a five-dimensional (5D) manifold foliated by a family of 4D hyper-surfaces, whose geometries are virtually indistinguishable from that of the 4D space-time of GR.  The axis of foliation is special in the sense that there could exist non-vanishing torsion components along that dimension. We impose  constraints on the connection so that any motion in the fifth dimension does not affect observations based on the geodesic motions along the 4D hyper-surfaces, thus keeping the fifth dimension essentially hidden.   The imposed constraints determine uniquely all the non-vanishing torsion components in terms of the 5D metric fields, making this a purely metric theory of gravity. Besides uniquely determining the torsion in the 5D geometry, the imposed constraints lead to interesting equivalence between the 5D geometry with torsion and the  torsion-free 4D geometry of GR. In particular, the components of the connection and the Ricci tensor along the 4D hypersurfaces turn out to exactly match what would arise from GR on a 4D space-time. Consequently, any test of this theory based on  geodesic motions will yield the same results as GR. 

Though by construction, the extra-dimension is hidden at the level of geodesic motion, its effect is clearly reflected in the field equations. The field equations are obtained by imposing the constraints on the action and varying it with respect to the metric. This  leads to global solutions qualitatively distinct from those obtained from GR.  Most notably, we find positive mass naked singularity solutions that match the Schwarzschild solution at large distances but lack an event horizon. In the cosmological context, we find oscillatory solutions in contrast to the inevitable singular big bang in standard cosmology.  

We begin  section II with a review of the general framework of the 5D geometry. Section III deals with the specification of the constraints and the determination of the torsion and connection in terms of the metric. Section IV is devoted to the derivation of modified Einstein equations from the standard action principle using Ricci scalar as the Lagrangian density. In section V, we apply the modified Einstein equations to the homogeneous and isotropic cosmology and identify numerical solutions pointing to accelerating and oscillatory solutions to the universe. In section VI, we discuss static spherically symmetric vacuum solutions and demonstrate the existence of positive mass naked singularity solutions. The final section is devoted to a summary and discussion of the results.

\section{General Framework of 5D geometry} 

\begin{figure}
\begin{center}
\includegraphics[scale=0.30]{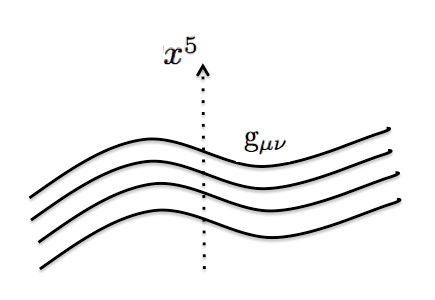}
\end{center}
\caption{Schematic representation of the 5D geometry.}
\label{fig:5Dschema}
\end{figure}

We denote the  coordinates of the 5D manifold by the latin indices, $i,j,k,...$ that take values 0,1,2,3 and 5, and the coordinates along the 4D hypersurfaces  by the greek indices, $\mu,\nu,\lambda,...$ that take values 0,1,2 and 3.    
Fig.~\ref{fig:5Dschema}  is a schematic representation of  the 5D geometry. With $x^{5}$ denoting the axis of foliation, the metric of the foliated 5D geometry has the form:
\begin{equation}
\mathbf{g}_{i j}=
\left[
\begin{array}{ccc|c}
\, & \, & \, & \, \\
\, & \mathrm{g}_{\mu \nu}+ \epsilon \mathrm{A}_{\mu} \mathrm{A}_{\nu} \Phi^{2} 
& \, & \,\,\, \epsilon \mathrm{A}_{\mu}\Phi^{2} \,\,\, \\ 
\, & \, & \, & \, \\
\hline
\, & \epsilon \mathrm{A}_{\nu}\Phi^{2}  
&\, & \epsilon \Phi^{2}  \\
\end{array}\right ]
\label{met1}
\end{equation}

\begin{equation}
\mathbf{g}^{i j}=
\left[
\begin{array}{ccc|c}
\, & \, & \, & \, \\
\, & \,\,\,\, \mathrm{g}^{\mu \nu} \,\,\,\, 
& \, & -\mathrm{A}^{\mu}  \\ 
\, & \, & \, & \, \\
\hline
\, & -\mathrm{A}^{\nu}  
&\, & \mathrm{A}_{\lambda} \mathrm{A}^{\lambda} +\epsilon  \Phi^{-2}  \\
\end{array}
\right ]
\label{met2}
\end{equation}

\begin{eqnarray}
\mathbf{g}_{\mu \nu} &=& \mathrm{g}_{\mu \nu}+\epsilon \A_{\mu}\A_{\nu}\Phi^{2},\,
\mathbf{g}_{\mu 5} = \epsilon \A_{\mu}\Phi^{2}, \,
\mathbf{g}_{5 5}=\epsilon \Phi^{2}, \nonumber \\
\mathbf{g}^{\mu \nu} &=& \mathrm{g}^{\mu \nu}, \,
 \mathbf{g}^{\mu 5}=-\A^{\mu}, \,
\mathbf{g}^{5 5} = \A_{\lambda} \A^{\lambda} + \epsilon \Phi^{-2} .
\label{eq:5Dmetric}
\end{eqnarray}
Here $\A_{\mu}$ is  a 4D vector, whose indices are raised and lowered with respect to the 4D metric $\mathrm{g}^{\mu \nu}$ and $\mathrm{g}_{\mu \nu}$. The fifth dimension is space-like if  $\epsilon= + 1$ and it is time-like if  $\epsilon=-1$.   
Note that the 5D metric is denoted by the bold face $\mathbf{g}$ and the 4D metric is light faced $\mathrm{g}$.

Let us denote the connection in the 5D geometry by $\Christilde{}{}{}$ and its antisymmetric part, the torsion by $\T{}{}{}$. 
\begin{equation}
\T{i}{j}{k}=\Christilde{i}{j}{k}-\Christilde{i}{k}{j}.
\end{equation}

Denoting the covariant derivative induced by the connection by $\tilde{\nabla}$, the metricity condition is expressed as $\tilde{\nabla}_{k}  \mathbf{g}_{i j} =0$. 
With the metricity condition, the connection  $\Christilde{}{}{}$ can be expressed as a sum of the Levi-Civita connection  $\Chrishat{}{}{}$ and the contorsion $\K{}{}{}$,
\begin{equation}
\Christilde{i}{j}{k} = \Chrishat{i}{j}{k} + \K{i}{j}{k},
\label{chris1}
\end{equation}

where the Levi-Civita connection is expressed purely in terms of the metric
\begin{equation}
\Chrishat{i}{j}{k}= \Big\{ {}_{j} {}^{i} {}_{k} \Big\}
=\frac{1}{2}\mathbf{g}^{im}[{\partial_{j}\mathbf{g}_{km}+\partial_{k}\mathbf{g}_{jm}-\partial_{m}\mathbf{g}_{jk}}],
\label{def:Chris}
\end{equation}
and the contorsion in terms of the torsion \cite{Nakaharabook}.
\begin{equation}
\K{i}{j}{k}= \frac{1}{2} \left[ \T{i}{j}{k} + \Ti{i}{j}{k} + \Ti{i}{k}{j} \right] .
\label{def:contorsion}
\end{equation}

In the absence of torsion, the connection is simply the Levi-Civita part. In order to compare the dynamics of this geometry to GR, we consider a reference space-time in four dimensions  with the metric $\g_{\mu \nu}$ and torsion-free 4D Levi-Civita connection $\Chris{\lambda}{\mu}{\nu}$. 
\begin{equation}
\Chris{\lambda}{\mu}{\nu}
=\frac{1}{2} \g^{\lambda \alpha}
  [{\partial_{\mu}\g_{\alpha \nu}+\partial_{\nu} \g_{\mu \alpha}-\partial_{\alpha}\g_{\mu \nu}}] .
\label{def:4DChris}
\end{equation}

This connection is different from the 4D components of the 5D Levi-Civita connection that contains additional terms $\Chrisdash{\lambda}{\mu}{\nu}$ (see eq.~\ref{eq:Chrishat} in the appendix) that depend on the extra-dimensional metric fields  $\A_{\mu}$ and $\Phi$. Hence the 4D components of the 5D Levi-Civita connection can be written as 
\begin{equation}
\Chrishat{\lambda}{\mu}{\nu} = \Chris{\lambda}{\mu}{\nu} + \Chrisdash{\lambda}{\mu}{\nu} 
\end{equation} 
In the presence of  torsion, with the inclusion of contorsion, the 4D components of the 5D connection takes the form 
\begin{equation}
\Christilde{\lambda}{\mu}{\nu} = \Chris{\lambda}{\mu}{\nu} + \Chrisdash{\lambda}{\mu}{\nu}  + \K{\lambda}{\mu}{\nu} .
\label{eq:extraterms}
\end{equation} 
We note that the additional terms $(\Chrisdash{\lambda}{\mu}{\nu}  + \K{\lambda}{\mu}{\nu})$ do not generally vanish. However, in the next section we impose constraints on the connection and find that these terms do vanish.

\section{Constraints on the Connection}  

With minimal modifications to standard GR in mind, we first assume that the 4D components of the connection $\Christilde{\lambda}{\mu}{\nu}$ are symmetric, that is (i) $\T{\lambda}{\mu}{\nu}=0$. Next, we require that geodesic motion and its observable effects in 4D are not affected by any motion in the fifth dimension. This requirement essentially ensures that the fifth dimension stays hidden at the level of 4D geodesics. For this purpose, considering  the 4D components of the geodesic equations in the 5D geometry, namely,
\begin{eqnarray}
&\overset{..}{x}^{\lambda}& +\Christilde{\lambda}{\mu}{\nu} \dot{x}^{\mu} \dot{x}^{\nu} 
+\left( \Christilde{\lambda}{\mu}{5}  + \Christilde{\lambda}{5}{\mu} \right) \dot{x}^{\mu} \dot{x}^{5}
+\Christilde{\lambda}{5}{5} \left( \dot{x}^{5} \right)^{2}=0, \nonumber
\label{eq:geodesic}
\end{eqnarray}
we are led to the second constraint (ii) $\Christilde{\lambda}{i}{5} =  \Christilde{\lambda}{5}{i} =0$.   An alternative formulation of these constraints in terms of vielbeins is worked out in \cite{MPLA}. These constraints are clearly not tensorial in nature because the fifth dimension is singled out. It turns out however, that they are sufficient to determine uniquely all the non-vanishing torsion components  in terms of the metric (see Appendix A for details).
\begin{eqnarray}
\T{\mu}{i}{j}&=& 0, \nonumber \\
\T{5}{\mu}{\nu} &=& 2 \partial_{[\mu} \A_{\nu]} +2 \J_{[\mu} \A_{\nu]},  \nonumber \\
\T{5}{\mu}{5} &=& \J_{\mu} - \partial_{5} \A_{\mu} -\A_{\mu} \J_{5}, 
\label{torsions}
\end{eqnarray}
where $\mathrm{J}_{i} \equiv  \Phi^{-1} \partial \Phi / \partial x^i$.

Using the above results for torsion and equations \ref{chris1}, \ref{def:Chris} and \ref{def:contorsion} we find the connection coefficients,
\begin{eqnarray}
\Christilde{\lambda}{5}{5}&=&\Christilde{\lambda}{\nu}{5}=\Christilde{\lambda}{5}{\nu}=0, \nonumber \\
\Christilde{5}{\mu}{\nu}&=&\nabla_{\mu}\A_{\nu}+ \J_{\mu} \A_{\nu}, \nonumber \\
\Christilde{5}{5}{\mu}&=& \partial_{5} \A_{\mu} + \J_{5} \A_{\mu}, \nonumber \\
\Christilde{5}{\mu}{5} &=& \J_{\mu}, \,\,\, \Christilde{5}{5}{5} =
\J_{5}, \,\,\,
\Christilde{\lambda}{\mu}{\nu}=\Chris{\lambda}{\mu}{\nu}.
\label{connections}
\end{eqnarray}
Here $\nabla_{\mu}$ is the  covariant derivative operator in the torsion free 4D geometry with metric $\g_{\mu \nu}$. This connection has a very special property: its 4D components are exactly the 4D Levi-Civita connection. That is, $\Chrisdash{\lambda}{\mu}{\nu}$ and $\K{\lambda}{\mu}{\nu}$ in eq.~\ref{eq:extraterms} exactly cancel each other.  

In addition to determining the  torsion and the connection in terms of the metric fields, the constraints also  imply that the 4D metric on all the  hypersurfaces are identical. 
As a consequence, the 4D components of the connection  also do not depend on $x^5$.  
\begin{equation}
\frac{\partial  \mathrm{g}_{\mu \nu} }{\partial x^5}=0 \,\,\,  \Longrightarrow \,\,\, 
\frac{\partial  \Christilde{\lambda}{\mu}{\nu} }{\partial x^5}=0.
\end{equation}
This should be contrasted with the Kaluza-Klein type theories where it is \emph{ a priori}  assumed that $\g_{\mu \nu}$, $\A_{\mu}$ and $\Phi$ are independent of $x^{5}$, known as the cylindrical condition. In our framework, though $\g_{\mu \nu}$ is required to be independent of $x^{5}$, $\A_{\mu}$  and $\Phi$ can in principle depend on $x^{5}$.

Substituting the connection (eq.~\ref{connections}) in the Ricci tensor defined by
\begin{equation}
\tilde{R}_{ik}=\partial_{k}\Christilde{j}{j}{i} - \partial_{j} \Christilde{j}{k}{i}
+\Christilde{j}{k}{m}\Christilde{m}{j}{i}- \Christilde{j}{j}{m}\Christilde{m}{k}{i},
\label{def:Ricci}
\end{equation}
we find
\begin{equation}
\tilde{R}_{\mu \nu}= R_{\mu \nu}, \,\, \tilde{R}_{\mu 5} = \tilde{R}_{5 \mu} = \tilde{R}_{5 5}=0. 
\label{defineRicci}
\end{equation}
Here $R_{\mu \nu}$  represents the Ricci tensor constructed from the torsion-free 4D Levi-Civita connection. Hence the 4D components of the Ricci tensor exactly match the Ricci tensor in GR with the metric $\g_{\mu \nu}$. 
It also follows that the 5D Ricci scalar is exactly the same as the Ricci scalar in the torsion-free 4D space-time,  that is $\tilde{R} =R$. 

An important point to emphasize is that at the level of geometry, this framework is virtually indistinguishable from the torsion-free 4D  space-time of GR. Any observable geodesic motion or geodesic deviations between  particles would match  what we expect based on GR. However, this is true only to the extent the metric $\g_{\mu \nu}$ is identical to the solution of the  Einstein's equations in GR. 
In section VI, we will see that this is  indeed  true in the weak field limit for spherically symmetric vacuum solutions  that are relevant for experimental observations within the solar system.

\section{Action principle and Modified Einstein Equations}

We start with the standard Einstein-Hilbert action with Ricci scalar as the Lagrangian density, 
\begin{equation}
S= \int \tilde{R} \sqrt{-\mathbf{g}} \, d^{5}x  .
\end{equation}
In varying the action, we note that the Ricci scalar and the connection coefficients described in the previous section are functions of the metric components alone. 
\begin{eqnarray}
 \delta S &=& \int \left[   \tilde{R} \, \delta \sqrt{-\mathbf{g}} 
+   \tilde{R}_{i k}   \, \delta \mathbf{g}^{ik} \, \sqrt{-\mathbf{g}}  
  \right] d^{5}x \nonumber \\
&+&  \int \delta \tilde{R}_{i k}\,\, \mathbf{g}^{ik} \sqrt{-\mathbf{g}}   \,\,  d^{5}x .
\label{eq:actionvary}
\end{eqnarray}
The first term gives rise to the usual Einstein tensor,
\[  
\tilde{\G}_{ik} = \tilde{R}_{i k} -(1/2) \mathbf{g}_{i k} \tilde{R}.
\]
In the absence of torsion, the second term becomes a boundary integral which vanishes when the variation is fixed at the boundary and hence will not contribute to the equations of motion. But in the presence of torsion, the second term gives a nonzero contribution. 
\begin{widetext}   

From eq.~\ref{def:Ricci}, we find the variations of the Ricci tensor  to be 
\begin{eqnarray}
\delta \tilde{R}_{ik}&=&\partial_{k} \delta \Christilde{j}{j}{i} - \partial_{j} \delta \Christilde{j}{k}{i}
+\Christilde{j}{k}{m} \delta \Christilde{m}{j}{i} +  \Christilde{m}{j}{i} \delta \Christilde{j}{k}{m}  
- \Christilde{j}{j}{m} \delta \Christilde{m}{k}{i} - \Christilde{m}{k}{i}  \delta \Christilde{j}{j}{m} \nonumber \\
&=&
\left[ \tilde{\nabla}_{k} \delta \Christilde{j}{j}{i} - \tilde{\nabla}_{j}  \delta \Christilde{j}{k}{i}  \right]
+ \T{m}{k}{j}  \delta \Christilde{j}{m}{i} .
\end{eqnarray} 
Then, the second term in the r.h.s of eq.~\ref{eq:actionvary} takes the form 
\begin{equation}
\int \delta \tilde{R}_{i k}  \, \mathbf{g}^{ik} \sqrt{-\mathbf{g}}  \, d^{5}x =
  \int \left[ \tilde{\nabla}_{k}  (\mathbf{g}^{ik} \delta  \Christilde{j}{j}{i} ) 
 - \tilde{\nabla}_{j} (\mathbf{g}^{ik}   \delta \Christilde{j}{k}{i} )  \right] \sqrt{-\mathbf{g}}  \, d^{5}x
+ \int \mathbf{g}^{ik}  \T{m}{k}{j}  \delta \Christilde{j}{m}{i} \sqrt{-\mathbf{g}}  \, d^{5}x  .
\label{eq:var2}
\end{equation}
In deriving the above equation, we have used the metricity condition, namely $\tilde{\nabla}_{j}\mathbf{g}^{ik}=0$.  Substituting for the covariant derivative in the first term of the r.h.s of  eq.~\ref{eq:var2},
\begin{eqnarray}
  \int  \left[ \tilde{\nabla}_{k}  (\mathbf{g}^{ik} \delta  \Christilde{j}{j}{i} ) \right.
 &-& \left. \tilde{\nabla}_{j} (\mathbf{g}^{ik}   \delta \Christilde{j}{k}{i} )  \right] \sqrt{-\mathbf{g}}  \, d^{5}x
 =\int \left[ \partial_{k}  (\mathbf{g}^{ik}  \delta \Christilde{j}{j}{i} \sqrt{-\mathbf{g}} \,) - \partial_{j} (\mathbf{g}^{ik}    \delta \Christilde{j}{k}{i} \sqrt{-\mathbf{g} }\,)  \right]  \, d^{5}x \\ \nonumber
&+& 
 \int   \mathbf{g}^{ik} \delta \Christilde{j}{j}{i} 
 \left[ \Christilde{m}{m}{k} -\frac{ \partial_{k}  \sqrt{-\mathbf{g}} }{ \sqrt{-\mathbf{g}} } \right] 
 \sqrt{-\mathbf{g}} \, d^{5}x
-  \int   \mathbf{g}^{ik} \delta \Christilde{j}{k}{i} 
\left[ \Christilde{m}{m}{j} - \frac{\partial_{j} \sqrt{-\mathbf{g}}}{\sqrt{-\mathbf{g}} }  \right] 
  \sqrt{-\mathbf{g}} \, d^{5}x   .
\end{eqnarray}
The first term in the r.h.s of the above equation is a boundary term, an integral of a total divergence. This will vanish when the variation is fixed at the boundary, and hence  can be ignored. The second and third terms in the r.h.s can be simplified by noting $\Christilde{m}{m}{k} = \Chrishat{m}{m}{k} + \T{m}{m}{k}$, and $ \Chrishat{m}{m}{k} = (\partial_{k}  \sqrt{-\mathbf{g}}) / \sqrt{-\mathbf{g}}$, leading to 
\begin{equation}
  \int \left[ \tilde{\nabla}_{k}  (\mathbf{g}^{ik} \delta  \Christilde{j}{j}{i} ) 
 - \tilde{\nabla}_{j} (\mathbf{g}^{ik}   \delta \Christilde{j}{k}{i} )  \right] \sqrt{-\mathbf{g}}  \, d^{5}x
 = -  \int \T{m}{k}{m}  \mathbf{g}^{ik} \delta \Christilde{j}{j}{i} \sqrt{-\mathbf{g}}  \, d^{5}x
+  \int \T{m}{j}{m}  \mathbf{g}^{ik} \delta \Christilde{j}{k}{i} \sqrt{-\mathbf{g}}  \, d^{5}x  ,
\end{equation}
and eq.~\ref{eq:var2} becomes
\begin{equation}
\int \delta \tilde{R}_{i k}  \, \mathbf{g}^{ik} \sqrt{-\mathbf{g}}  \, d^{5}x =
 \int  \left[ - \T{m}{k}{m} \mathbf{g}^{ik}  \delta \Christilde{j}{j}{i}  
+  \T{m}{j}{m} \mathbf{g}^{ik}    \delta \Christilde{j}{k}{i} 
+ \T{m}{k}{j}  \mathbf{g}^{ik}  \delta \Christilde{j}{m}{i} \right] \sqrt{-\mathbf{g}}  \, d^{5}x  .
\label{eq:varyfinal}
\end{equation}

\end{widetext}

The second term in the r.h.s of eq.\ref{eq:actionvary} is thus a function of torsion and variations in the connection given by eq.~\ref{eq:varyfinal}. Clearly, this term vanishes if we assume that torsion is zero, and action principle would yield the standard Einstein equations.  Alternatively, if we treat the variation in the connection to be composed of independent variations in metric and torsion, we obtain the Einstein-Cartan equations \cite{Hehl}, which  ultimately leads to zero torsion when the  matter is not coupled to the connection. However, since torsion is not an independent degree of freedom in our framework and is a function of the metric components given by eq.~\ref{torsions}, we first substitute its components in terms of the metric and then carry out the variation with respect to the metric. To this end, we note,
\begin{eqnarray}
\T{m}{k}{m} \mathbf{g}^{i k} \delta \Christilde{j}{j}{i} &=& 
 \T{5}{\nu}{5}  [ \mathbf{g}^{\mu \nu} (\delta \Christilde{\alpha}{\alpha}{\mu}  + \delta \Christilde{5}{5}{\mu} )
+ \mathbf{g}^{5 \nu}   \delta \Christilde{5}{5}{5}] ,
\nonumber \\
\T{m}{j}{m} \mathbf{g}^{i k} \delta \Christilde{j}{k}{i} &=& 
\T{5}{\nu}{5} \mathbf{g}^{\mu \alpha} \delta \Christilde{\nu}{\mu}{\alpha} ,
\nonumber \\
\T{m}{k}{j}  \mathbf{g}^{ik}  \delta \Christilde{j}{m}{i} &=& 
\T{5}{\nu}{5}  [ \mathbf{g}^{\mu \nu} \delta \Christilde{5}{5}{\mu} +
\mathbf{g}^{5 \nu}   \delta \Christilde{5}{5}{5}] .
\end{eqnarray}
Taking these terms together, eq.~\ref{eq:varyfinal} takes the form
\begin{eqnarray}
&& \int \delta \tilde{R}_{i k}  \, \mathbf{g}^{ik} \sqrt{-\mathbf{g}}  \, d^{5}x =  \nonumber \\
&& \int   \T{5}{\nu}{5} 
\left[ \g^{\mu \alpha} \delta \Chris{\nu}{\mu}{\alpha} - \ g^{\mu \nu} \delta \Chris{\alpha}{\alpha}{\mu} \right]
\sqrt{-\mathbf{g}}  \, d^{5}x ,
\label{pre-extraterm}
\end{eqnarray}
Note that the variations in the connection, $\delta \Chris{\nu}{\mu}{\alpha}$ and $\delta \Chris{\alpha}{\alpha}{\mu}$,  involve only the 4D components. Since these are the 4D Levi-Civita components that only depend on the 4D metric $\g_{\mu \nu}$, the above equation takes the form
\begin{equation}
 \int \delta \tilde{R}_{i k}  \, \mathbf{g}^{ik} \sqrt{-\mathbf{g}}  \, d^{5}x =
  \int \h_{\mu \nu} \delta \g^{\mu \nu} \, \sqrt{-\mathbf{g}} \, d^{5}x ,
\label{extraterm}
\end{equation}   
where (see Appendix B for details),
\begin{eqnarray}
&&
\h_{\mu \nu} = \nabla_{( \mu} \B_{\nu) } - (\nabla \cdot \B ) \g_{\mu \nu}  + \J_{( \mu} \B_{\nu )} - (\J \cdot \B) \g_{\mu \nu} ,
  \nonumber \\
&&
 \qquad \B_{\mu}  \equiv \T{5}{\mu}{5} = \J_{\mu} - \partial_{5} \A_{\mu} -\A_{\mu} \J_{5}.
\label{eq:defH}
\end{eqnarray}

Taking together the variations in both terms in eq.~\ref{eq:actionvary}, we obtain the modified Einstein tensor. 
\begin{eqnarray}
\tilde{\G}_{\mu \nu} &=& R_{\mu \nu} - \frac{1}{2} ( \g_{\mu \nu} + \A_{\mu} \A_{\nu} \epsilon \Phi^{2} ) R  + \h_{\mu \nu}
= \Sigma_{\mu \nu} ,
\nonumber \\
\tilde{\G}_{\mu 5} &=& - \frac{1}{2} \A_{\mu} \epsilon \Phi^{2} R = \Sigma_{\mu 5} ,
\nonumber \\
\tilde{\G}_{5 5} &=& - \frac{1}{2} \epsilon \Phi^{2} R = \Sigma_{5 5} , \nonumber
\end{eqnarray} 
where  $\Sigma$ is the stress energy tensor that arises from the variations of the assumed matter fields in the Lagrangian. Our focus being on gravity, we will not further discuss the origin of $\Sigma$.  
Since the physical interpretation of the stress- energy is more transparent with one covariant and one contravariant indices, we express the above equations in an alternate form, by noting $\tilde{\G}_{i}^ {\,\, j} = \mathbf{g}^{k j} \tilde{G}_{i k}$,
\begin{eqnarray}
R_{\mu}^{\,\, \nu} -\frac{1}{2} R \delta^{\,\, \nu}_{\mu} + \h^{\,\, \nu}_{\mu} &=& \Sigma_{\mu}^{\,\, \nu},
\label{EE1} \\
-\A^{\alpha} R_{\mu \alpha} - \A^{\alpha} \h_{\mu \alpha} &=& \Sigma_{\mu}^{\,\, 5},
\label{EE2} \\
0 = \Sigma_{5}^{\,\, \mu},  \qquad  -\frac{1}{2} R &=& \Sigma_{5}^{\,\, 5} .
 \label{EE3}
\end{eqnarray}
These are the modified Einstein equations in our framework. 
Since by construction the fifth dimension is hidden with respect to the observable 4D motion, the 5D components of the stress tensor $\Sigma_{\mu}^{\,\, 5}$ and $\Sigma_{5}^{\,\, 5}$ are unobservable. It is not possible to solve equations \ref{EE2} and \ref{EE3} unless these components are theoretically known from the  5D matter Lagrangian. In the present formulation, for simplicity, we shall ignore these equations as though they simply serve to evaluate the components $\Sigma_{\mu}^{\,\, 5}$ and $\Sigma_{5}^{\,\, 5}$, and treat only eq.~\ref{EE1} with the observable 4D stress tensor to be relevant to physical solutions. In the absence of specified matter fields in the Lagrangian, an alternate way to interpret the modified Einstein equations is to regard  $-\h^{\,\, \nu}_{\mu}$  as \emph{extra-dimensionally induced matter}. 

When $\h^{\,\, \nu}_{\mu}=0$, eq.~\ref{EE1} reduces to the standard Einstein equations for the 4D metric components $\g_{\mu \nu}$.  In this case the 4D Bianchi identity necessarily implies the conservation of matter $\nabla_{\nu}  \Sigma^{\,\, \nu}_{\mu} =0$.  But in general when  $\h^{\,\, \nu}_{\mu}$ is non-vanishing and dependent on the extra-dimensional metric fields $\A_{\mu}$ and $\Phi$, eq.~\ref{EE1} by itself may not be sufficient to solve for  $\g_{\mu \nu}$ along with $\A_{\mu}$ and $\Phi$, even after fixing the gauge.  However, an important  physical simplification  can be achieved by generalizing the cylindrical condition to assume that $\A_{\mu}$ and $\Phi$ do not depend on  $x^{5}$. With this assumption, $\B_{\mu}=\J_{\mu}$ and $\h^{\,\, \nu}_{\mu}$ depends only on $\Phi$ and not on $\A_{\mu}$, and eq.~\ref{EE1} is sufficient to solve for both $\g_{\mu \nu}$ and $\Phi$. The vector $\A_{\mu}$ can in principle be evaluated from eq.~\ref{EE2} by setting $\Sigma_{\mu}^{\,\, 5}$ to zero, but this would be inconsequential as $\A_{\mu}$ is decoupled from the physically relevant equation  that solves for the 4D metric $\g_{\mu \nu}$. Hence, in the rest of the paper we will make the assumption of cylindrical condition in order to explore solutions of physical interest to the modified Einstein equations.   

Finally, when  $\h^{\,\, \nu}_{\mu}$ is non-vanishing, we note that $ \Sigma^{\,\, \nu}_{\mu}$ does not necessarily have to satisfy the 4D matter conservation. However, with minimum modifications to GR and the empirical conservation laws in mind, it is reasonable to assert the conservation of $ \Sigma^{\,\, \nu}_{\mu}$. Since the standard Einstein tensor satisfies the 4D Bianchi identity independently of  $\h^{\,\, \nu}_{\mu}$, the 4D matter conservation implies,   
\begin{equation}
\nabla_{\nu}  \Sigma^{\,\, \nu}_{\mu} =0  \Longrightarrow 
\nabla_{\nu}  \h^{\,\, \nu}_{\mu} =0. 
\label{conservation}
\end{equation}

In the reminder of the paper, we study the solutions to the modified Einstein equations (eq.~\ref{EE1})  in  two extremely symmetric situations, namely, the homogeneous-isotropic geometry and the static spherically symmetric geometry. 

\section{Homogeneous-Isotropic Cosmology}

The 4D metric of a homogeneous and isotropic universe has the form
\begin{equation}
ds^{2} = - dt^{2} +a^{2}(t) \left( \frac{dr^2}{1-kr^2} +r^2 d\Omega^{2}  \right)
\end{equation}
The values of $k=0,+1,-1$  correspond respectively to flat, closed, and hyperbolic spatial geometries.   The standard Einstien tensor for this metric is given by \cite{Waldbook}  
\begin{eqnarray}
\mathrm{G}^{t}_{t} &=& 3 (\dot{a}/a)^{2}  + 3 k/a^2 , \nonumber \\
\mathrm{G}^{r}_{r} &=&  2 (\overset{..}{a}/a) + (\dot{a}/a)^2 + k/a^2 , \nonumber \\
 \mathrm{G}^{\theta}_{\theta} &=& \mathrm{G}^{\phi}_{\phi} = \mathrm{G}^{r}_{r} ,
 \label{CosmG}
\end{eqnarray}
where  over-dot denotes a derivative with respect to time. Since the geometry is spatially homogeneous and isotropic, the metric fields including $\A_{\mu}$ and $\Phi$ in the 5D geometry  only depends on time. Hence the only non-vanishing component of $\J_{\mu}$ is  $\J_{t}$. The induced matter terms given in  eq.~\ref{eq:defH} are
\begin{eqnarray}
\h^{\,\, t}_{t}  &=& 3\J_{t} (\dot{a}/a) ,  \nonumber \\
 \h^{\,\, r}_{r} &=& 2 \J_{t} (\dot{a}/a) +  \dot{\J}_{t} +  \J^{2}_{t} , \nonumber \\
 \h^{\,\, \theta}_{\theta}  &=& \h^{\,\, \phi}_{\phi} = \h^{\,\, r}_{r} .
\end{eqnarray}

\begin{figure*}[!ht]
\begin{center}
\includegraphics[scale=0.5]{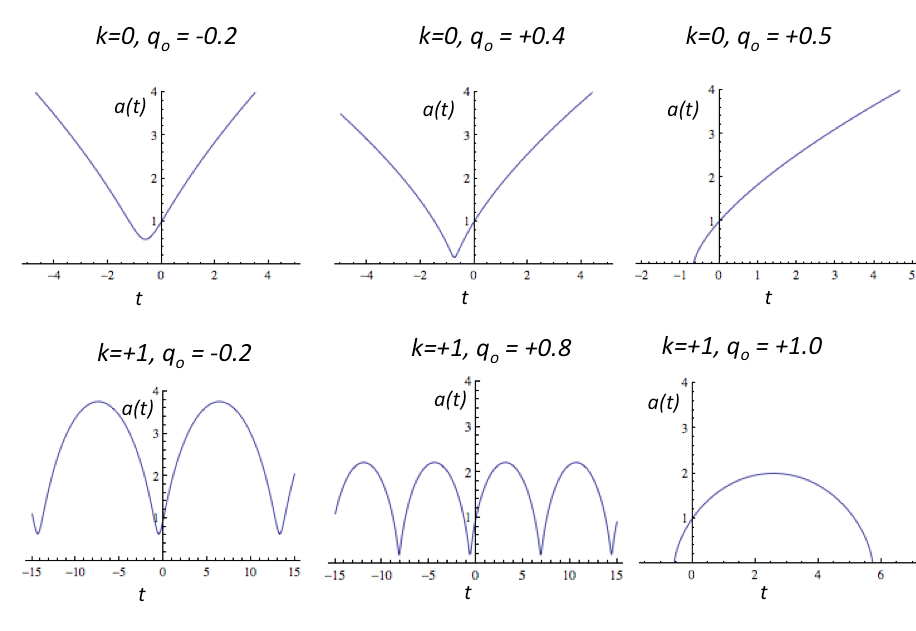}
\caption{Solutions for $a(t)$ for different values of deceleration parameter for $k=0$ and $k=+1$.} 
\label{Cosmologyfig}
\end{center}
\end{figure*}

Before writing out the modified Einstein equations, we note that the conservation equation (eq.~\ref{conservation}) now reduces to 
\begin{eqnarray}
\nabla_{\nu} \h^{\,\, \nu}_{\mu} =   3 \J_{t} \left[  (\overset{..}{a}/a) - (\dot{a}/a) \J_{t} \right] =0  \nonumber \\
\nonumber \\ 
\Longrightarrow \qquad \J_{t} =0 \qquad \mathrm{or} \qquad  \J_{t} =  \overset{..}{a}/\dot{a}
\end{eqnarray}

 These are the only two possibilities. From the definition of $\J_{\mu}$, this implies that either $\Phi$ is a constant, which would give rise to the usual Friedman-Robertson-Walker cosmology, or $\Phi= \dot{a}(t)$. Focusing on the latter case, $\h^{\,\, \nu}_{\mu}$ simplifies to 
\begin{equation}
\h^{\,\, t}_{t} = 3 \overset{..}{a}/a, \qquad  
\h^{\,\, r}_{r} =  2 (\overset{..}{a}/a)  +  (\overset{...}{a}/\dot{a}).
\label{CosmH}
\end{equation}

 Taking the stress tensor  to be that of a perfect fluid, the modified Einstein equations (eq.~\ref{EE1}) take the form
\begin{eqnarray}
 3 (\dot{a}/a)^{2}  + 3 k/a^2 + 3 \overset{..}{a}/a  &=&  8 \pi \rho 
  \label{FRW}  \\
 4 (\overset{..}{a}/a)   + (\dot{a}/a)^2 + (\overset{...}{a}/\dot{a})  + k/a^2  &=& -  8 \pi P 
  \end{eqnarray}
where $\rho$ and $P$ are the density and pressure of the 4D matter. Combining the above equations, we find   
\begin{equation}
\dot{\rho} + 3(\rho + P) \dot{a}/a =0 ,
\label{cons}
\end{equation}  
which is just a restatement of the 4D matter conservation equation. 
For matter dominated universe, $P=0$, and consequently eq.~\ref{cons} yields
 \begin{equation}
 \rho a^{3} =constant  \qquad \mathrm{or} \qquad \rho = \rho_o /a^3 .
 \label{cons_matter}
 \end{equation}
 In effect,   eqns.~\ref{FRW} and \ref{cons_matter} are sufficient to solve for $a(t)$, which needs two initial conditions along with the specification of $\rho_o$. 
 \begin{equation}
 a^2 \overset{..}{a} + a (\dot{a})^2 + k a =  8 \pi \rho_o /3  ,
\end{equation} 
 Without loss of generality we choose the  current epoch to be at $t=0$, set the current size of the universe $a(0)=1$, and the unit of time such that $\dot{a}(0)=1$. In this unit of time, the current value of the Hubble's constant will be 1.  Rather than specifying the value of $\rho_o$, we choose to specify the current value of  $\overset{..}{a}$. The effective equation for $a(t)$ then takes the form
 \begin{equation}
 a^2 \overset{..}{a} + a (\dot{a})^2 + k a = 1+k-q_o  ,
\end{equation}
where $q_o =  - a(0) \overset{..}{a}(0) / \dot{a}^{2}(0) $, the current value of the deceleration parameter, is the only free parameter to be specified. Figure \ref{Cosmologyfig} shows the behavior of $a(t)$ for various values of $q_o$ for spatially flat and closed topologies. For the spatially flat topology,  shown in the top panels of the figure, we find that the universe does not originate from a singular big bang for all  $q_o<+0.5$. For the spatially closed topology shown in the bottom panels, we find oscillatory solutions for all $q_o<+1$.  

Oscillatory solutions in the spatially closed topology exhibit a scale factor  that oscillates between a maximum $a_{max}$ and a minimum $a_{min}$. The acceleration reaches a positive value in a narrow interval around $a_{min}$, and then becomes negative for the rest of the cycle until it gets back near $a_{min}$.   By taking the value of $q_o$ arbitrarily close to 1, we can make $a_{min}$ arbitrarily close to zero. This can be seen from the bottom-left and bottom-middle panels of fig.~\ref{Cosmologyfig}.  Thus one could construct a universe that collapses and bounces back to expand when it reaches an arbitrarily small size, or equivalently arbitrarily high energy densities. 
It remains to be seen if such solutions would fit the empirical red shift data.

\section{Static Spherically Symmetric Vacuum Solutions} 

The most general static spherically symmetric 4D metric  has the form\footnote{ The scalar functions
$A(r)$ and $B(r)$ defined in this section should not be confused with the vectors $\A_{\mu}$ and $\B_{\mu}$ defined in eqns. \ref{eq:5Dmetric} and \ref{eq:defH} respectively.}

\begin{equation}
ds^{2} = - A(r) dt^{2} + B(r) dr^{2} + r^{2} d \Omega^{2} ,
\end{equation}
and the standard Einstein tensor for this metric is \cite{Waldbook} 
\begin{eqnarray}
\mathrm{G}^{t}_{t} &=& \frac{rB' + B^2 -B }{r^2 B^2}, \nonumber \\
\mathrm{G}^{r}_{r} &=&\frac{AB - rA'-A}{r^2 AB} , \nonumber \\
\mathrm{G}^{\theta}_{\theta} &=& \frac{2A^2 B' - 2AB A' -2rAB A'' + rB A'^2 +rA A'B'}{4rA^2 B^2} ,\nonumber \\
\mathrm{G}^{\phi}_{\phi} &=& \mathrm{G}^{\theta}_{\theta} ,
\end{eqnarray}
where a prime in the above equations denotes a derivative  with respect to $r$. 

The additional term $\h^{\,\, \nu}_{\mu}$ in the modified Einstein equations (eq.~\ref{EE1}) depends only on $\Phi$ when the cylindrical condition is imposed on all metric components and is given by
\begin{equation}
\h^{\,\, \nu}_{\mu} = \nabla_{ \mu} \J^{\nu } - (\nabla \cdot \J ) \delta_{\mu}^{\nu}  + \J_{ \mu} \J^{\nu } - (\J \cdot \J) \delta_{\mu}^{\nu} .
\end{equation}
Since $\J_{\mu}=\Phi^{-1} \partial_\mu \Phi$, the quantity $\nabla_{ \mu} \J^{\nu }$ is intrinsically symmetric in $\mu$ and $\nu$. The static spherical symmetry of the geometry implies  that $\J_{r}$ is the only non-vanishing component, which we denote by $J(r)$. With this,
\begin{eqnarray}
\h^{\,\, t}_{t} &=& \frac{rJB'  -2B\left(rJ'+2J+rJ^2 \right) }{2r B^2} , \nonumber \\
\h^{\,\, r}_{r} &=&\frac{-J \left(rA' +4A\right)}{2r AB} ,  \nonumber \\
\h^{\,\, \theta}_{\theta} &=& \frac{rJA B' - 2rAB J'-rJB A' -2ABJ \left(1+rJ \right)}{2rA B^2} , \nonumber \\
\h^{\,\, \phi}_{\phi}&=& \h^{\,\, \theta}_{\theta}.
\end{eqnarray}

In order to obtain vacuum solutions,  we set $\Sigma_{\mu}^{\nu}=0$ in eq.~\ref{EE1} and find the following three equations.
\begin{eqnarray}
J' &=& - \frac{J(1+B)}{r} ,  \nonumber  \\
A' &=& - \frac{2A(1-B+2rJ)}{r(2+Jr)} ,  \nonumber \\
B' &=&  \frac{2B \left( r^2J^2+ (1+rJ)(1-B) \right)}{r (2+Jr)} .
\label{eq:BHEE}
\end{eqnarray}
A close examination of the above equations reveals two  basic properties of the function $J(r)$. First,  if $J(r)$ is a constant, it has to be identically zero. 
Secondly,  if $J(r)$ vanishes at some point, it has to  vanish identically everywhere. 
The simplest solution to the coupled equations (eq.~\ref{eq:BHEE})  is when $J(r)$ vanishes everywhere, 
\begin{equation}
J(r)=0  , \, A(r)= \left(1-\frac{2M}{r} \right)  , \,B(r) = \left(1-\frac{2M}{r} \right)^{-1}  ,
\end{equation}
which of course is the well-known Schwarzschild solution as expected.

\subsection{General solution to $J(r)$}

Let $F(r) \equiv 1/rJ(r)$ when $J(r)$ is non-vanishing. Substituting for $J(r)$ in terms of $F(r)$,  the coupled equations (eq.~\ref{eq:BHEE}) lead to the following equation for $F(r)$,
\begin{equation}
r \left[(2F^2+F)F'' +F'^2\right] =  F' (F+2), 
\label{Feq}
\end{equation} 
whose solution in turn determines the 4D metric functions $A(r)$ and $B(r)$.
The obvious solution of eq.~\ref{Feq} is  $F(r)=$constant. This leads to  
\begin{equation}
J(r)=c/r \, , \, A(r)= (r)^{- \frac{2+4c}{2+c} }  \, , \,B(r) = 0 ,
\end{equation}  
which is clearly  unacceptable because $B(r)$ is identically zero. 

Assuming that $F(r)$ is not a constant, we can obtain solutions to the second order differential equation (eq.~\ref{Feq}). In principle, the solution would have two integration constants that would be determined by the boundary conditions, one of which immediately follows from the form of the equation.  It can be easily  seen that if $F(r)$ is a solution, then $F(\lambda r)$ is also a solution for any scaling constant $\lambda$. We find a general solution in the implicit form 

\begin{equation}
\lambda^2 r^2 = \gamma^2  \left| F/\gamma + \beta +1 \right|^{1+\beta}  
\left| F/\gamma + \beta - 1 \right|^{1-\beta} ,
\label{Fsol2}
\end{equation}
where $\beta$ and $\gamma$ are defined in terms of an independent arbitrary constant $c$.
\begin{equation}
\gamma = \sqrt{1+c+c^2}, \,\, \,
\beta = \frac{1+c}{\sqrt{1+c+c^2}}. \, \,
\end{equation}

\begin{figure}
\begin{center}
\includegraphics[scale=0.30]{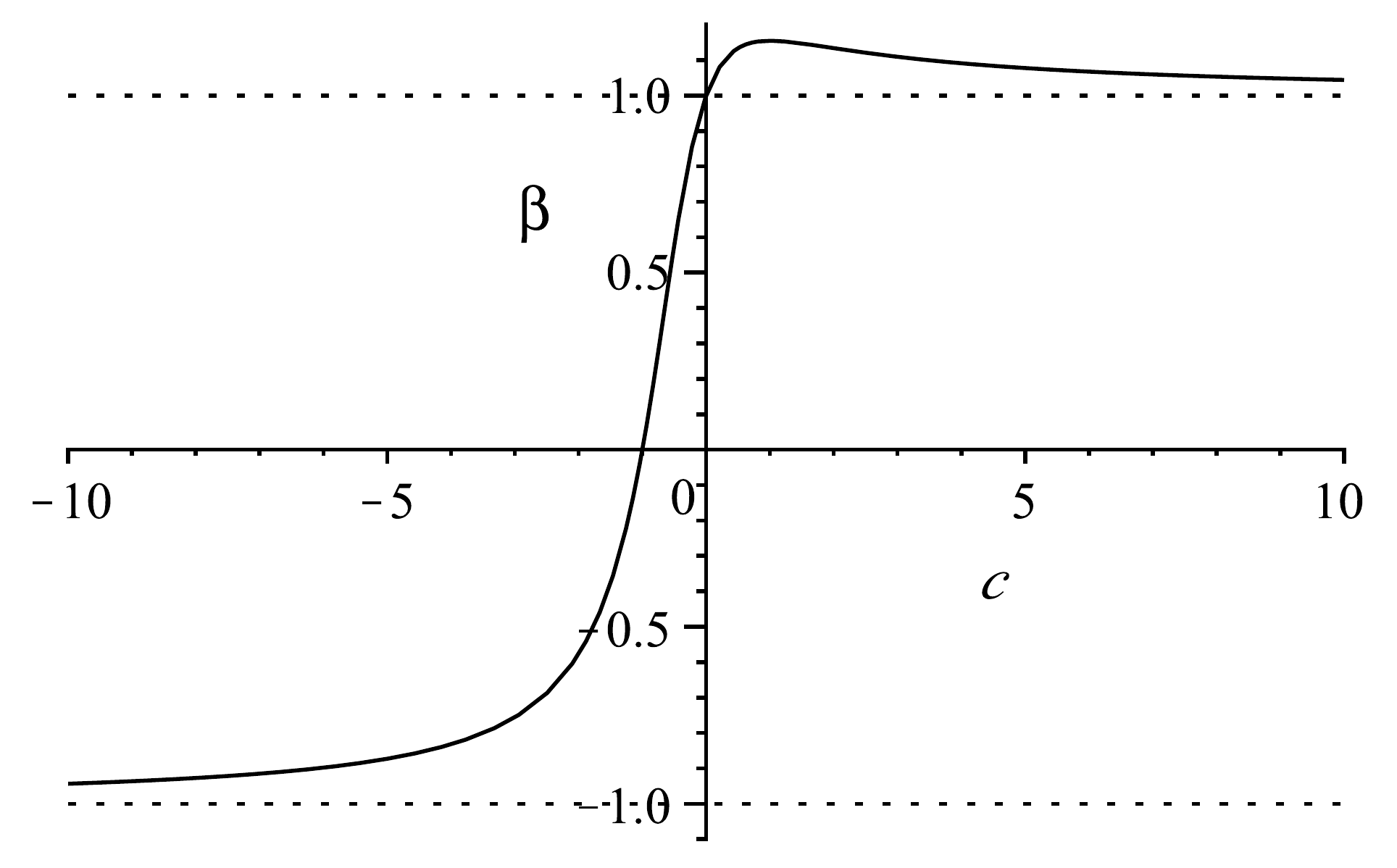}
\end{center}
\caption{$\beta$ is plotted as a function of $c$. }
\label{fig:beta}
\end{figure}

With $\lambda$ and $c$ as two arbitrary constants, eq.~\ref{Fsol2} represents the general solution to the second order  differential equation (eq.~\ref{Feq}). From eq.~\ref{Fsol2}, we find the derivatives of $F(r)$ to be 
\begin{equation}
F' = \frac{1}{r} \frac{F^2 +2(1+c)F + c}{F},  \qquad F'' = -\frac{c}{r} \frac{F'}{F^2},
\label{Fderiv}
\end{equation}
and substituting them in eq.~\ref{eq:BHEE}, we obtain the metric functions $A(r)$ and $B(r)$ in terms of $F(r)$. 
\begin{eqnarray}
A'(r) &=& 2cA(r)/rF(r)  ,  \label{eq:feqns} \\
B(r) &=&  1+2(1+c)/F(r)+ c/F^2(r)   .
\label{eq:feqns1}   
\end{eqnarray}

In order to obtain asymptotically flat solutions, we shall impose the boundary conditions $A(r \rightarrow \infty)=1$ and $B(r \rightarrow \infty)=1$. 
To understand the behavior of the functions  $A(r)$ and $B(r)$ which define the observable 4D geometry, we start with  the properties of $F(r)$.

\begin{figure*}[!ht]
\begin{center}
\includegraphics[scale=0.45]{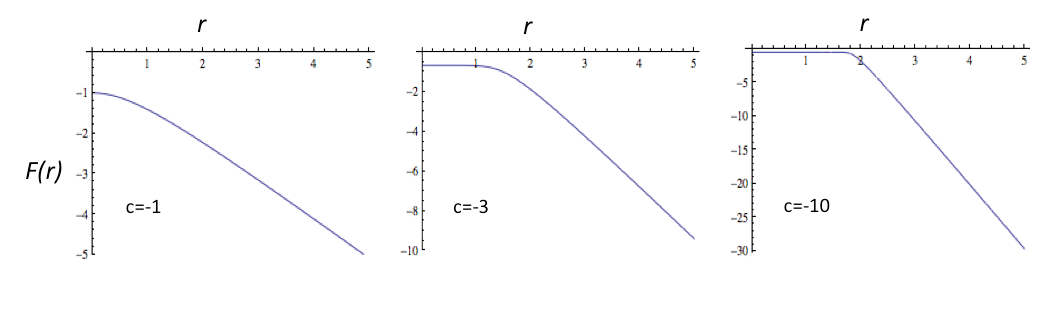}
\caption{Numerical solution to $F(r)$  with $\lambda=-1$. } 
\label{fig:Fsol}
\end{center}
\end{figure*}

Unfortunately eq. \ref{Fsol2} does not yield an explicit functional form for $F(r)$ except for simple cases when $c$ is either 0 or -1. Nevertheless the relevant properties of $F(r)$ can be inferred from analyzing this implicit function. First note that the quantity $(1+c+c^2)$ is positive definite and $\beta$ is finite and bounded for all values of $c$.   Figure.~\ref{fig:beta} plots the behavior of $\beta$ to show that it asymptotically reaches +1 and -1 at $c= +\infty$ and $-\infty$ respectively. The following observations summarize the qualitative properties of $F(r)$.

1) At $r =0$, $F$ can take one of two possible values.  If $c<0$ ($\beta <1$), then $F(0)$ can be either $\gamma (-\beta-1)$ or $\gamma(-\beta+1)$, while if $c>0$ ($\beta>1$), then $F(0)$ can only be $\gamma(-\beta-1)$.

2) In the limit $r \rightarrow \infty$, $F$ necessarily has to diverge in order to satisfy the boundary condition $B(r \rightarrow \infty)=1$. Eq.~\ref{Fsol2} then implies for large $r$, $\lambda^2 r^2 =  F^2$, implying $F$ could be either positive or negative, such that  
\begin{equation}
F(r\rightarrow \infty) = \lambda r  \Rightarrow J(r\rightarrow \infty) =  \lambda^{-1}/r^2 .
\end{equation}
The behavior of $F(r)$ at the extremities is summarized in the following table.
\begin{center}
  \begin{tabular}{ | c  || c | c | }
    \hline 
     & $c<0 \,\,\, (\beta<1)$ & $c>0 \,\,\,(\beta>1) $ \\ \hline \hline
    $r =0$ & $F = \begin{cases} \gamma(-\beta +1) \, >0 \\  \gamma(-\beta-1) \, <0  \end{cases}$  
             & $F = \gamma(- \beta -1) \, < 0$ \\  \hline
    $r \rightarrow \infty$ & $F  =  \lambda r  $  &  $F  = \lambda r  $    \\
    \hline
  \end{tabular}
  \label{tab}  
\end{center}

3) From eq.~\ref{Fderiv}, it can be shown that $F'=0$ when $F$ is either $\gamma(-\beta+1)$ or $\gamma(-\beta-1)$, which can happen only at $r=0$. Hence $F'$ is either positive definite or negative definite, and so $F(r)$ is a monotonic function.

4) Though $F$ is monotonic and finite for any finite $r$, it can reach zero at $r_o$ given by  
\begin{equation}
\lambda^2 r_o^2 = |c| \left| \frac{\beta +1}{\beta -1}\right|  ^{\beta}, 
\end{equation}
and from eq.~\ref{Fderiv}, $F'$ diverges at $r_o$. Considering the physical relevance  of these solutions, we shall only  focus on solutions that are non-vanishing everywhere. Such solutions do indeed exist for a range of parameter values. Rewriting eq.~\ref{Fsol2} at $r=r_o$, 
\begin{equation}
\left| \frac{F(r_o)}{\gamma (\beta +1)} +1 \right|^{1+\beta} 
\left| \frac{F(r_o)}{\gamma (\beta -1)} +1 \right|^{1-\beta}  =1 ,
\end{equation}
we note that $F(r_o)=0$ is not the only solution. Numerical plots in fig.~\ref{fig:Fsol} demonstrates the existence of non-vanishing $F(r)$ solutions. 

5) From the table above, (i) for $c>0$, since $F(0)$ is negative, $F(r)$ has to be negative definite which requires $\lambda$ to be negative. (ii) For $c<0$, $\lambda$ can be either positive or negative, making $F(r)$ either positive definite or negative definite respectively.  
   
The functional form of $F(r)$ described by the above five properties  along with  equations \ref{eq:feqns} and \ref{eq:feqns1} will  yield the functional form of the metric functions $A(r)$ and $B(r)$.   
   
 \vspace{20mm}

\subsection{Metric functions $A(r)$ and $B(r)$ }

With the boundary condition $A(\infty) =B(\infty) =1$, eqns.~\ref{eq:feqns} and \ref{eq:feqns1} yield 
\begin{equation}
A(r)  = \exp{ \left( - \int_{r}^{\infty} \frac{2c}{rF(r)} dr \right)} ,
\label{Aeqn}
\end{equation}
\begin{equation}
B(r) = 1+ 2(1+c)/F(r) + c/F^2(r) .
\label{Beqn}
\end{equation}
The following observations summarize the qualitative behavior of $A(r)$ and $B(r)$. 

1) From the asymptotic behavior of $F(r) \rightarrow \lambda r$ for large $r$, we note that   
\begin{eqnarray}
A(r) &=& 1 - \frac{2c \lambda^{-1} }{r} +   \mathcal{O}(1/r^2) , \\
B(r) &=& 1 + \frac{2(1+c) \lambda^{-1} }{r} +   \mathcal{O}(1/r^2) .
\end{eqnarray}
Hence, when $|c| \gg 1$ and $r \rightarrow \infty$, the above solutions approximate the Schwarzschild solution with mass $M \equiv [c \lambda^{-1}] $. When both $c$ and $\lambda$ are either positive or negative, the gravity is attractive,  while when one is positive and the other is negative, the gravity is repulsive. 

 2) Since $F(r)$ is either positive definite or negative definite, both $A(r)$ and $B(r)$ are finite and positive for all $r>0$.  At $r=0$, since $F(0)$ is either $\gamma(-\beta-1)$ or $\gamma(-\beta+1)$, eq.~\ref{Beqn} implies $B(0)=0$.  

3) As $r \rightarrow 0$, the integral in eq.~\ref{Aeqn} diverges as $[2c/F(0)] \ln(r)$. When $[c/F(0)]$ is positive, then $A(0)=0$, and when $[c/F(0)]$ is negative $A(0) = \infty$.  The sign of $[c/F(0)]$ is the same as the sign of $M=[c\lambda^{-1}]$. For $M>0$, $A(r)$ monotonically increases from $A(0)=0$  to $A(\infty)=1$; for $M<0$,  $A(r)$ monotonically decreases from $A(0)= +\infty$ to $A(\infty)= 1$. 

4) Irrespective of the sign of $M$, $B(0)=0$ and $B(\infty)=1$. However, $B(r)$ is not necessarily monotonic.  From eq.~\ref{Beqn}, we see that $B'=0$ when either $F'=0$ or when $F(r) = -c/(1+c)$. From the previous subsection, $F' \neq 0$ for all $r>0$, but $F(r)$ could attain the value $-c/(1+c)$ for certain values of $c$ and $\lambda$. 
Since $F(r)$ is a non-vanishing monotonic function taking all values from $F(0)$ to $\pm \infty$, it is straightforward to check if it would attain the value $-c/(1+c)$.  When $\lambda<0$, $F(r)$ is negative definite, and $-c/(1+c)$ needs to be a negative number lesser than $F(0)=\gamma(-\beta-1)$, which happens only when $c<-1$.  When $\lambda>0$, $F(r)$ is positive definite, and $-c/(1+c)$ needs to be a positive number greater than $F(0)=\gamma(-\beta+1)$, which happens only when $-1<c<0$.  

\begin{widetext}
\begin{figure*}[!ht]
\begin{center}
\includegraphics[scale=0.45]{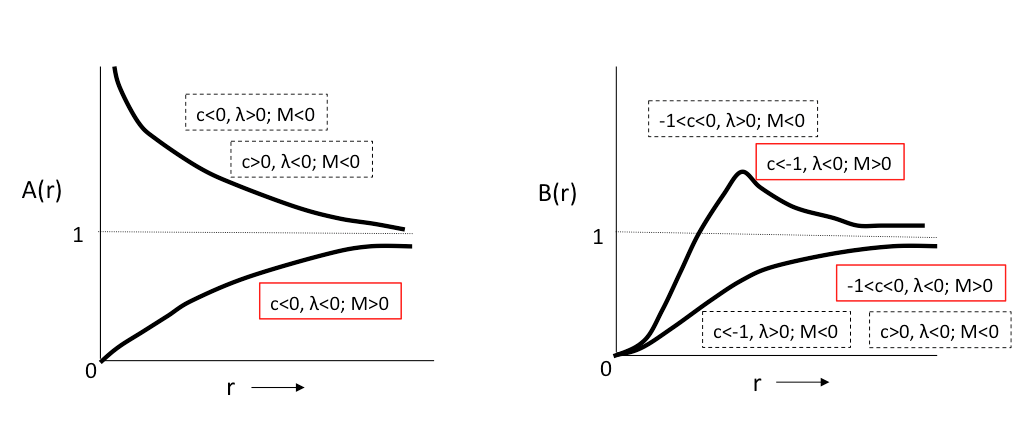}
\caption{Schematic behavior of  $A(r)$ and $B(r)$  } 
\label{fig:ABfig}
\end{center}
\end{figure*}
\end{widetext}

The qualitative behavior of the functions $A(r)$ and $B(r)$ for the various allowed ranges of $c$ and $\lambda$ are shown in figure \ref{fig:ABfig}.  When $M=[c\lambda^{-1}]$ is positive, $A(r)$ is a monotonically increasing function leading to attractive gravity, and corresponds to Schwarzschild solution at large $r$ when $|c| \gg1$.

\vspace{100mm}

\subsection{Naked singularity at $r=0$}

An important point to note from figure \ref{fig:ABfig} is that these solutions do not have an event horizon because both $A(r)$ and $B(r)$ are finite and positive for all $r>0$. 

Clearly, these solutions are smooth for all $r>0$. However, the point $r=0$ is a physical singularity.  Explicit calculations show that the Ricci scalar $R_{\mu}^{\mu}$ vanishes everywhere, but the quantity  $R_{\mu \nu} R^{\mu \nu} $ is nonvanishing.  It turns out that 
\begin{equation}
R_{\mu \nu} R^{\mu \nu} = \frac{2(3F^2(r) +2cF(r)+c^2)}{F^4(r) B^2(r) r^4}  .
\end{equation}
At $r=0$, the numerator does not generally vanish, but the denominator vanishes, making $R_{\mu \nu} R^{\mu \nu}$ diverge. Hence these solutions correspond to a naked singularity at $r=0$ with no event horizon to censor it.

\section{Summary and Discussion}

Metric and torsion are two independent constituents of metric compatible Riemannian geometry.  Because of the immense successes of torsion-free GR, torsion has not played a significant role in theories of gravity. However, when gravity is to be included with other interactions of elementary particles with intrinsic spin, a more general theory  including torsion becomes imperative \cite{Hehl}.

In the present work, torsion is incorporated in a novel way in higher dimensional Kaluza-Klein type theories. Here torsion is not an independent degree of freedom coupled to spin, rather it is determined in terms of metric through a set of physically motivated constraints, which serve  (i) to confine torsion to the extra dimension, leaving the 4D space-time torsion free, and  (ii) to ensure that geodesic motions in 4D remain totally unaffected by the presence of the extra-dimension. These constraints have previously been imposed in terms of  veilbeins \cite{MPLA, Viet}, but here it is realized that they impose essentially the requirement that the fifth dimension is hidden at the level of geodesic motion. It turns out that the non-vanishing torsion components are functions of the 5D metric components with the 4D metric $\g_{\mu \nu}$ obeying the so called cylindrical condition, namely, it is independent of $x^5$.  In the resulting geometry, all the 4D hypersurfaces are equivalent, and the 4D components of the  connection and the Ricci tensor exactly match those of  the standard 4D GR. Hence, at the level of geodesics, this geometry is virtually indistinguishable from that of the standard GR. 

To proceed further, we derive modified Einstein equations from the action principle with Ricci scalar as the Lagrangian density. In this respect, an alternate approach presents itself. In the Palatini formulation of GR, the action is varied independently with respect to the connection, and in the absence of torsion, the metricity follows from the equations of motion. Recently \cite{Sotiriou, DadhichPons}, it has been shown that even without assuming the absence of torsion, variations of the action independently with respect to the metric and the connection lead to GR equations along with metricity, provided the matter Lagrangian is not coupled to the connection. In our case, with the constraints imposed on the connection, it is more convenient and natural to impose metricity prior to action variation. Since the entire connection is determined to be a function of the metric, we only need to vary  the action with respect to the metric to obtain the modified equations, making the theory  a purely metric theory of gravity. However, adopting a Palatini-style approach, one could relax the assumption of metricity and vary the action independently with respect to metric and  connection along with the imposed constraints, which might lead to a different set of modified Einstein equations.

We apply the ensuing modified Einstein equations to study the cosmology of a homogenous-isotropic universe.  In matter dominated phase of the universe (zero pressure), we obtain a second order differential equation for the scale factor $a(t)$ in contrast to the first order differential equation in the usual FRW cosmology. In FRW cosmology, the second derivative of $a(t)$ cannot be independently prescribed as an initial condition, and decelerating expansion is a necessary outcome in the absence of a cosmological constant. However in our case, we have the choice of an initial condition for  the second derivative of $a(t)$, which can be tuned to fit the observed acceleration of the universe. Figure \ref{Cosmologyfig} presents the behavior of $a(t)$ for various choices of the current acceleration.

In an earlier version of the present work \cite{oldPaper}, the field equations were derived differently;  by varying the action with respect to the metric prior to expressing the Ricci tensor in terms of the metric. In the cosmology equations  generated from those field equations, acceleration was not an independently prescribable initial condition. Chen and Jing \cite{ChenJing} showed that those  equations  yield accelerating universe solutions without resorting to dark energy. They show that the model not only fits the supernovae data, but also solves the cosmic age problem of old high redshift objects  \cite{hiZshift}. Whether the cosmological solutions described in the current work would fit empirical results just as well needs to be investigated. 

In the case of spherically symmetric vacuum solutions to the modified Einstein equations, we find some remarkably interesting results. As is well known, in the unique Schwarzschild solution of GR, when the mass is positive, an event horizon censors the central singularity. In contrast, we find  positive mass naked singularity solutions that lack an event horizon. Recently, similar positive mass solutions without a horizon have been found \cite{KalyanaRama} in a simpler setting  of torsion-free GR with multiple extra-dimensions. It would be interesting to see how the particular vacuum solutions in the torsion induced geometry in the present case match with those solutions in the torsion free geometry. 

The existence of positive mass solutions with naked singularity have immediate consequences on gravitational collapse, opening up the possibility of an arbitrarily large star collapsing to an arbitrarily small non-singular state. Since trapped surfaces would not necessarily form in such collapses, finite matter pressure could be sufficient to withstand a total collapse to singularity. This suggests a  detailed analysis of such solutions by treating $-\h^{\,\, \nu}_{\mu}$ in eq.~\ref{EE1} as extra-dimensionally induced matter in standard GR. It then raises the possibility of a sufficiently strong gravitational collapse that stops short of collapsing to a singularity with a finite induced stress-energy tensor that potentially violates the weak energy condition in the region near the center. It remains to be seen if an arbitrarily small static model star with finite stress-energy tensor can be constructed with the external geometry matching the type of solutions discussed in this paper.

In conclusion, inclusion of torsion in the context of extra-dimensions presents a novel way of obtaining modified Einstein equations that have significant physical consequences. For clarity and simplicity, we have confined the treatment to five dimensions. However, the framework can be generalized to arbitrary dimensions $D$, producing torsion-free  $D-1$ dimensional metric theory.  We could also consider many extra-dimensions and generalize the constraints so as to  hide  all the extra-dimensions using torsion, leading to a more general theory.

\section*{Acknowledgments}
The authors thank Venky Krishnan, Ramesh Anishetty, Johannes Noller and Stanley Deser for helpful discussions.

\newpage

\begin{widetext}

\appendix

\section{  Computing the geometric quantities in the 5D manifold }
In this appendix, expression for torsion, connection coefficients and the Ricci tensor of the 5D geometry are computed  in terms of the metric.  It will be shown that the physical constraints imposed on the connection will be sufficient to determine uniquely all the components of torsion and hence the other geometric quantities in terms of the metric.

\subsection{5D Levi-Civita Connections} \label{5DLevi}
The 5D Levi-Civita connection is given by  
\begin{equation}
\Chrishat{j}{i}{k}=\frac{1}{2}\mathbf{g}^{jm}\left( \partial_k \mathbf{g}_{im}+\partial_i \mathbf{g}_{km}
 -\partial_m \mathbf{g}_{ik} \right)
\end{equation}
Expressing the 5D metric in  terms of the 4D metric and the extra-dimensional metric fields given by eq.~\ref{eq:5Dmetric}, we find the 5D Levi-Civita connection to be
\begin{eqnarray}
\Chrishat{5}{\mu}{\nu}&=& \frac{1}{2}\left(\nabla_{\mu}\A_{\nu}+\nabla_{\nu}\A_{\mu}\right)
+\left( \A_{\mu} \J_{\nu} + \A_{\nu} \J_{\mu} \right)
+\frac{1}{2}\epsilon \Phi^{2}\A^{\lambda}\left(\A_{\mu}\F_{\lambda \nu}+\A_{\nu}\F_{\lambda \mu} \right) +\epsilon \Phi^{2}\A_{\mu}\A_{\nu} \A^{\lambda}\J_{\lambda} \nonumber \\
&-&\frac{1}{2} \left( \A^{\lambda} \A_{\lambda} +  \epsilon \Phi^{-2} \right) 
   \partial_{5}\left( \epsilon \A_{\mu}\A_{\nu} \Phi^{2}\right) 
-\frac{1}{2} \left( \A^{\lambda} \A_{\lambda} +  \epsilon \Phi^{-2} \right)  \partial_{5} \mathrm{g}_{\mu \nu} 
\\
\Chrishat{\mu}{5}{5} &=& -\epsilon \Phi^{2} \J^{\mu} + \epsilon \Phi^{2} \partial_{5}\A^{\mu} +\epsilon \Phi^{2} \A^{\mu} \J_{5}
-\epsilon \Phi^{2} \A_{\sigma} \partial_{5} \mathrm{g}^{\mu \sigma}  \label{Gmu55}
\\
\Chrishat{5}{5}{5}  &=& \epsilon \Phi^{2}\A^{\lambda}\left( \J_{\lambda} -\partial_{5} \A_{\lambda}\right) 
- \epsilon \Phi^{2} \A^{\lambda}\A_{\lambda} \J_{5} + \J_{5}
\\
\Chrishat{\mu}{5}{\nu}&=&\frac{1}{2}\epsilon \Phi^{2} \F_{\nu}^{\cdot \mu} 
- \epsilon \Phi^{2} \A_{\nu} \J^{\mu} 
+ \frac{1}{2} \mathrm{g}^{\mu \sigma} \partial_{5} \left( \epsilon \Phi^{2} \A_{\sigma} \A_{\nu} \right) 
+\frac{1}{2} \mathrm{g}^{\mu \sigma} \partial_{5} \mathrm{g}_{\nu \sigma}
\\
\Chrishat{5}{5}{\nu}&=&\frac{1}{2}\epsilon \Phi^{2} \A^{\lambda} \F_{\lambda \nu} +\epsilon \Phi^{2} \A_{\nu} \A^{\lambda} \J_{\lambda} + \J_{\nu} -\frac{1}{2}\A^{\lambda} \partial_{5} \left( \epsilon \A_{\lambda} \A_{\nu} \Phi^{2} \right) - \frac{1}{2} \A^{\lambda} \partial_{5} \mathrm{g}_{\nu \lambda}
\\
\Chrishat{\lambda}{\mu}{\nu}&=&\Chris{\lambda}{\mu}{\nu} + \Chrisbar{\lambda}{\mu}{\nu} 
\nonumber \\
 \Chrisbar{\lambda}{\mu}{\nu} &=&
\frac{1}{2}\epsilon \Phi^{2}\left(\A_{\nu} \F_{\mu}^{\cdot \lambda} +\A_{\mu} \F_{\nu}^{\cdot \lambda} \right) - \epsilon \Phi^{2} \A_{\mu} \A_{\nu}\J^{\lambda} +\frac{1}{2} \A^{\lambda} \partial_{5} \left( \epsilon \A_{\mu} \A_{\nu} \Phi^{2} \right) + \frac{1}{2} \A^{\lambda} \partial_{5} \mathrm{g}_{\mu \nu} \label{eq:Chrishat}
\end{eqnarray}
 where $\F_{\mu \nu} \equiv \partial_{\mu}\A_{\nu}-\partial_{\nu} \A_{\mu}$, $\J_{i} \equiv \epsilon \Phi^{-1}\partial_i \epsilon \Phi$ and  $\Chris{\lambda}{\mu}{\nu}$ is the 4D Levi-Civita connection obtained from the metric $\mathrm{g}_{\mu \nu}$. and $\nabla_{\mu}$ is the derivative operator with the 4D Levi-Civita connection. The raising and lowering of indices on $\F_{\mu \nu}$, $\A_{\mu}$ and $\J_{\mu}$ are performed with respect to the 4D metric.

\subsection{ Torsion components}

We start with the conditions that $\T{\lambda}{\mu}{\nu}=0$ and $\Christilde{\lambda}{i}{5}= \Christilde{\lambda}{5}{i}=0$.  Together these conditions imply  $\T{\lambda}{i}{j}=0$. The remaining non-vanishing components of torsion that need to be determinned are  $\T{5}{i}{j}$, a total of 10 independent components. The conditions  $\Christilde{\lambda}{5}{i}=0$, consisting of 20 equations are sufficient to determine  uniquely  all the non-vanishing torsion components. 
\begin{equation}
\Christilde{\mu}{i}{5} =   \Chrishat{\mu}{i}{5} + \K{\mu}{i}{5} =0
\label{appAcond}
\end{equation} 
From the above equation, the non vanishing components of contorsion and torsion can be determined in terms of components of the components of 5D Levi-Civita connection $\Chrishat{}{}{}$. First taking $i=5$,
\begin{equation}
\K{\mu}{5}{5} = \mathbf{g}^{\mu j} \mathbf{g }_{ 55} \T{5}{j}{5} 
= \mathbf{g}^{\mu \nu} \mathbf{g }_{ 55} \T{5}{\nu}{5}  = -\Chrishat{\mu}{5}{5} 
\end{equation}
From the 5D metric (eq.~\ref{met2}), we have $\mathbf{g}^{\mu\nu}=\mathrm{g}^{\mu\nu}$. Multiplying both sides by $\mathrm{g}_{\mu\sigma}$  and using the orthogonality relations of the metric,
\begin{eqnarray}   
\mathrm{g}_{\mu\sigma} \mathrm{g}^{\mu\nu} \mathbf{g }_{55} \T{5}{\nu}{5}&=&  -\mathrm{g}_{\mu\sigma} \Chrishat{\mu}{5}{5}\nonumber \\
 \mathbf{g }_{55} \T{5}{\sigma}{5}&=&  -\mathrm{g}_{\mu\sigma} \Chrishat{\mu}{5}{5}\nonumber \\
 \Rightarrow \T{5}{\sigma}{5} &=& -\mathrm{g}_{\mu\sigma} \Chrishat{\mu}{5}{5} \epsilon \Phi^{-2}
 \label{T5sigma5}
\end{eqnarray}
Substituting for $\Chrishat{\mu}{5}{5}$ from eq.~\ref{Gmu55}, we obtain
\begin{equation}
\T{5}{\sigma}{5} = \mathrm{g}_{\mu \sigma} [  \J^{\mu} - \partial_{5}\A^{\mu} - \A^{\mu} \J_{5}
+ \A_{\sigma} \partial_{5} \mathrm{g}^{\mu \sigma} ]
\label{eq:T5sigma5}
\end{equation}

We have thus far used 4 equations and solved for 4 of the 10 independent torsion components. Next, take $i=\nu$ in eq.~\ref{appAcond}. The contorsion components $\K{\mu}{\nu}{5}$ can be expressed in terms of the torsion components as follows: 
\begin{eqnarray}
\K{\mu}{\nu}{5} &=& \frac{1}{2} \mathbf{g}^{\mu j} \left(\mathbf{g}_{5 5} \T{5}{j}{\nu} +   \mathbf{g}_{\nu 5} \T{5}{j}{5}   \right)   \nonumber \\
&=& \frac{1}{2}  \mathbf{g}^{\mu 5} \left( \mathbf{g}_{5 5}  \T{5}{5}{\nu} \right) +\frac{1}{2}  \mathbf{g}^{\mu \sigma} 
\left( \mathbf{g}_{5 5} \T{5}{\sigma}{\nu} +   \mathbf{g}_{\nu 5} \T{5}{\sigma}{5}  \right) \nonumber \\
&=&  \frac{1}{2} \epsilon \A^{\mu}  \Phi^{2}  \T{5}{\nu}{5} +  \frac{1}{2} \epsilon \A_{\nu} \Phi^{2} \mathrm{g}^{\mu \sigma} \T{5}{\sigma}{5}
+ \frac{1}{2}\epsilon \Phi^{2} \mathrm{g}^{\mu \sigma} \T{5}{\sigma}{\nu}
\end{eqnarray}
From  $\K{\mu}{\nu}{5} =-\Chrishat{\mu}{\nu}{5}$, and using eq.~\ref{T5sigma5},  we have 
\begin{equation}
-2\Chrishat{\mu}{\nu}{5} + \A^{\mu} \mathrm{g}_{\alpha \nu} \Chrishat{\alpha}{5}{5} + \A_{\nu} \Chrishat{\mu}{5}{5} = \epsilon \Phi^{2} \mathrm{g}^{\mu \sigma }\T{5}{\sigma}{\nu}
\end{equation}
and hence,
\begin{equation}
\T{5}{\sigma}{\nu}=\epsilon \Phi^{-2}\mathrm{g}_{\mu \sigma } 
\left[ -2\Chrishat{\mu}{\nu}{5}+\A^{\mu} \mathrm{g}_{\lambda \nu}\Chrishat{\lambda}{5}{5}+\A_{\nu} \Chrishat{\mu}{5}{5} \right]
\label{T5sigmanu}
\end{equation} 
These are 16 equations. Though the torsion in the l.h.s is antisymmetric, the r.h.s is a combination of symmetric and antisymmetric terms. Substituting for the 5D Levi-Civita connection from section~(\ref{5DLevi}), the  above equation has the form,
\begin{eqnarray}
\T{5}{\sigma}{\nu}&=&  \left[ \partial_{\sigma} \A_{\nu} - \partial_{\nu} \A_{\sigma}  \right]
+ \left[ \J_{\sigma} \A_{\nu} - \J_{\nu} \A_{\sigma} \right]  \nonumber \\
 & +& \left[ \epsilon \Phi^{-2} \partial_{5} \mathrm{g}_{\sigma \nu}
+\A_{\sigma} \A^{\lambda} \partial_{5} \mathrm{g}_{\lambda \nu} 
+\A_{\nu} \A^{\lambda} \partial_{5} \mathrm{g}_{\lambda \sigma} \right] 
\end{eqnarray}
The first two terms in the r.h.s above are antisymmetric in $\sigma$ and $\nu$, while the third term is  symmetric. The antisymmetry of torsion implies that the symmetric terms in the r.h.s must be zero, and hence
\begin{equation}
\partial_{5} \mathrm{g}_{\sigma \nu} =0.
\label{eq:d5=0}
\end{equation}
Consequently,
\begin{equation}
\T{5}{\sigma}{\nu}=  \left[ \partial_{\sigma} \A_{\nu} - \partial_{\nu} \A_{\sigma}  \right]
+ \left[ \J_{\sigma} \A_{\nu} - \J_{\nu} \A_{\sigma} \right] .
\label{eq:T5sigmanu}
\end{equation}

Thus the 20 equations of imposed condition (eq.~\ref{appAcond}) have determined all 10 independent non-vanishing components of the torsion (equations \ref{eq:T5sigma5} and \ref{eq:T5sigmanu}), and in addition imposed a constraint on the 10 independent components of the 4D metric $\mathrm{g}_{\sigma \nu}$ making them independent of $x^5$.

\subsection{The contorsion, connection and the Ricci tensor}

Some components of contorsion are directly prescribed by the imposed condition, namely $\K{\mu}{i}{5} =\K{\mu}{5}{i} =-\Chrishat{\mu}{i}{5} $.  The remaining components of the contorsion can be calculated from the torsion components by using  eq.~\ref{def:contorsion}. Since $\T{5}{i}{j}$ are the only non vanishing components of torsion, it follows that 
\begin{eqnarray}
\K{\mu}{i}{5} &=& \K{\mu}{5}{i} = -\Chrishat{\mu}{i}{5},   \nonumber \\
\K{5}{i}{5}&=&   \frac{1}{2} \T{5}{i}{5} + \frac{1}{2} \mathbf{g}^{5 j} 
        \left(\mathbf{g}_{i 5} \T{5}{j}{5} + \mathbf{g}_{5 5} \T{5}{j}{i}   \right) \nonumber \\
\K{5}{5}{i}&=&\K{5}{i}{5} + \T{5}{5}{i} \nonumber \\
\K{\mu}{\nu}{\lambda} &=&\frac{1}{2} \mathbf{g}^{\mu j} \left(\mathbf{g}_{\nu 5} \T{5}{j}{\lambda}     
      +\mathbf{g}_{\lambda 5} \T{5}{j}{\nu}   \right)   \nonumber \\
\K{5}{\nu}{\lambda}&=&  \frac{1}{2} \T{5}{\nu}{\lambda}+ \frac{1}{2} \mathbf{g}^{5 j} 
        \left(\mathbf{g}_{\nu 5} \T{5}{j}{\lambda} + \mathbf{g}_{\lambda 5} \T{5}{j}{\nu}   \right) 
\end{eqnarray}
Substituting for the nonvanishing torsion components given by equations \ref{eq:T5sigma5} and \ref{eq:T5sigmanu}, along with the requirement that the 4D metric is independent of the fifth dimension (eq.~\ref{eq:d5=0}), the components of contorsion are found to be
\begin{eqnarray}
\K{\mu}{i}{5} &=& \K{\mu}{5}{i} = -\Chrishat{\mu}{i}{5},  
 \nonumber \\
\K{5}{i}{5}&=& - \Chrishat{5}{i}{5} + \mathrm{J}_{i}, 
\qquad  
\K{5}{5}{i}=\K{5}{i}{5} + \T{5}{5}{i} ,
\nonumber \\
\K{\mu}{\nu}{\lambda} &=&\epsilon \Phi^{2} \F^{\mu}_{\cdot (\lambda} \A_{\nu)}
+\A_{\lambda} \A_{\nu}\J^{\mu} \epsilon \Phi^{2} - \frac{1}{2} \A^{\mu} \partial_{5}( \epsilon \A_{\nu} \A_{\lambda} \Phi^2) ,
  \nonumber \\
\K{5}{\nu}{\lambda}&=& - \epsilon \Phi^{2}  \A^{\sigma} \F_{ \sigma (\lambda}  \A_{\nu)} 
- (\A_{\sigma}\J^{\sigma}) \epsilon \A_{\nu} \A_{\lambda} \Phi^{2}
+\frac{1}{2}(\A_{\sigma}\A^{\sigma} +\epsilon \Phi^{-2}) \partial_5 (\epsilon \A_{\nu} \A_{\lambda} \Phi^2 ) -\A_{\nu} \J_{\lambda}  
+\frac{1}{2} \F_{\nu \lambda}.
\label{contorsions}
\end{eqnarray}

Now from eq.~\ref{chris1}, we obtain all the connection coefficients, 
\begin{eqnarray}
\Christilde{5}{\mu}{\nu}&=&\nabla_{\mu}\A_{\nu}+ \J_{\mu} \A_{\nu}, \nonumber \\
\Christilde{5}{5}{\mu}&=& \partial_{5} \A_{\mu} + \J_{5} \A_{\mu}, \nonumber \\
\Christilde{5}{\mu}{5} &=& \J_{\mu},  \,\, \Christilde{5}{5}{5}= \J_{5} \nonumber \\
\Christilde{\lambda}{\mu}{\nu}&=&\Chris{\lambda}{\mu}{\nu},  \nonumber \\
\Christilde{\mu}{5}{5}&=&\Christilde{\mu}{\nu}{5}=\Christilde{\mu}{5}{\nu}=0 .
\end{eqnarray}

Taking $i=\mu$ and $k=\nu$ in the Ricci tensor defined by eq.~\ref{defineRicci},  we have 
\begin{equation}
\begin{array}{cccc}
\tilde{R}_{\mu \nu} = & + \partial_{\nu}\Christilde{\sigma}{\sigma}{\mu} - \partial_{\sigma} \Christilde{\sigma}{\nu}{\mu}
& +\Christilde{\sigma}{\nu}{\lambda}\Christilde{\lambda}{\sigma}{\mu} 
&- \Christilde{\sigma}{\sigma}{\lambda}\Christilde{\lambda}{\nu}{\mu} \\
\, & + \partial_{\nu}\Christilde{5}{5}{\mu} - \partial_{5} \Christilde{5}{\nu}{\mu} 
\, &+\Christilde{5}{\nu}{\lambda}\Christilde{\lambda}{5}{\mu} 
&- \Christilde{5}{5}{\lambda}\Christilde{\lambda}{\nu}{\mu}  \\
\, & \, & +\Christilde{\sigma}{\nu}{5}\Christilde{5}{\sigma}{\mu} 
&- \Christilde{\sigma}{\sigma}{5}\Christilde{5}{\nu}{\mu}  \\ 
& \, & +\Christilde{5}{\nu}{5}\Christilde{5}{5}{\mu} 
&- \Christilde{5}{5}{5}\Christilde{5}{\nu}{\mu}
\end{array}
\label{Chriseq}
\end{equation}
Since $\Christilde{}{}{}$ is the same as $\Chris{}{}{}$ when all the indices are four dimensional,  the first line is clearly the 4D Ricci tensor. The terms in the subsequent lines can be re-expressed in terms of the 4D covariant derivative operator as follows  
\begin{equation}
\tilde{R}_{\mu \nu} = R_{\mu \nu} + \nabla_{\nu} \Christilde{5}{5}{\mu} -\partial_{5} \Christilde{5}{\nu}{\mu}
-\Christilde{5}{5}{5} \Christilde{5}{\nu}{\mu} + \Christilde{5}{\nu}{5} \Christilde{5}{5}{\mu} 
\end{equation}
Substituting for the connection $\Christilde{}{}{}$ from eq.~\ref{Chriseq}, we find after some algebra this can be simplified as
\begin{eqnarray}
\tilde{R}_{\mu \nu} &=& R_{\mu \nu} +
\nabla_{\nu}(\partial_5 \A_{\mu} +\J_{5}\A_{\mu}) -\partial_{5}(\nabla_{\nu}\A_{\mu}+\J_{\nu}\A_{\mu})
-\J_{5}(\nabla_{\nu}\A_{\mu}+\J_{\nu}\A_{\mu})+\J_{\nu}(\partial_5 \A_{\mu}+\J_{5}\A_{\mu}) \nonumber \\
&=& R_{\mu \nu}+ \A_{\lambda} \partial_{5} \Chris{\lambda}{\mu}{\nu}  \nonumber \\
&=& R_{\mu \nu}
\end{eqnarray}
Similarly, the other components of the Ricci tensor are found to be 
\begin{eqnarray}
\tilde{R}_{\mu 5} &=&  \partial_{5}\Christilde{\sigma}{\sigma}{\mu} =0  \\
\tilde{R}_{5 \nu} &=& \partial_{\nu} \Christilde{5}{5}{5} -\partial_{5} \Christilde{5}{\nu}{5}=0 \\
\tilde{R}_{5 5} &=&0
\end{eqnarray}

Note that neither the connection nor the Ricci tensor depend on the signature ($\epsilon$) of the fifth dimension.

\newpage

\section{ Computing the modified Einstein tensor }

Here we provide some intermediate steps to go from equations~\ref{pre-extraterm} and \ref{extraterm} to eq.~\ref{eq:defH} and obtain a simplified expression for $\h_{\mu \nu}$. With $\B_{\nu} \equiv \T{5}{\nu}{5}$, and $\h_{\mu \nu}$ defined by
\begin{equation}
 \int   \B_{\nu}  \left[ \g^{\mu \alpha} \delta \Chris{\nu}{\mu}{\alpha} -  \g^{\mu \nu} \delta \Chris{\alpha}{\alpha}{\mu} \right]
\sqrt{-\mathbf{g}}  \, d^{5}x  
= \int \h_{\mu \nu} \delta \g^{\mu \nu} \, \sqrt{-\mathbf{g}} \, d^{5}x ,
\label{eq:defH2}
\end{equation}
we show that 
\begin{equation}
\h_{\mu \nu} = \nabla_{( \mu} \B_{\nu) } - (\nabla \cdot \B ) \g_{\mu \nu}  + \J_{( \mu} \B_{\nu )} - (\J \cdot \B) \g_{\mu \nu} .
\label{eq:defH3}
\end{equation}

\textbf{Proof:}

Consider the first term in the integrand of the l.h.s of eq.~\ref{eq:defH2}.
\begin{equation}
2 \g^{\mu \alpha} \delta  \Chris{\nu}{\mu}{\alpha} =
 \g^{\mu \alpha} \g^{\nu \lambda} (\partial_{\mu} \delta \g_{\lambda \alpha} + \partial_{\alpha} \delta \g_{\lambda \mu} - \partial_{\lambda} \delta \g_{\mu \alpha} ) +
\g^{\mu \alpha}  (\partial_{\mu} \g_{\lambda \alpha} + \partial_{\alpha} \g_{\lambda \mu} - \partial_{\lambda} \g_{\mu \alpha} ) \delta \g^{\nu \lambda} .
\end{equation}
The variations of the covariant metric in the above equation can be re-expressed in terms of the variations of the contravariant metric using the identity $\delta \g_{\alpha \lambda} = - \g_{\alpha \mu} \g_{\lambda \nu} \delta \g^{\mu \nu}$.  We note that 
\begin{eqnarray}
\g^{ \mu \alpha} \g^{\nu \lambda} \partial_{\mu} \delta \g_{\lambda \alpha} 
&=&
-  \g^{\nu \lambda} \partial_{\mu} \g_{\lambda \sigma} (\delta \g^{\sigma \mu}) 
-  \g^{\mu \alpha}  \partial_{\mu}  \g_{\beta \alpha} ( \delta \g^{\nu \beta} )
-  \partial_{\mu} (\delta \g^{\nu \mu} )      \nonumber \\ 
\g^{\mu \alpha} \g^{\nu \lambda} \partial_{\alpha} \delta \g_{\lambda \mu} 
&=& \g^{\mu \alpha} \g^{\nu \lambda} \partial_{\mu} \delta \g_{\lambda \alpha}   \nonumber \\ 
-\g^{\mu \alpha} \g^{\nu \lambda} \partial_{\lambda} \delta \g_{\mu \alpha} 
&=& 
  \g^{\nu \lambda} \partial_{\lambda} \g_{\mu \sigma} (\delta \g^{\sigma \mu} ) 
+ \g^{\nu \lambda}  \partial_{\lambda}  \g_{\beta \sigma} ( \delta \g^{\sigma \beta} )
+ \g^{\nu \lambda}  \g_{\beta \sigma} \partial_{\lambda} ( \delta  \g^{\sigma \beta} )  \nonumber  
\end{eqnarray}
Taken together, we obtain
\begin{equation}
2 \g^{\mu \alpha} \delta  \Chris{\nu}{\mu}{\alpha}= -  2\g^{\nu \lambda} \partial_{\mu} \g_{\lambda \sigma} (\delta \g^{\sigma \mu}) -  2 \partial_{\mu} (\delta \g^{\nu \mu} ) 
 +2 \g^{\nu \lambda} \partial_{\lambda} \g_{\mu \sigma} (\delta \g^{\sigma \mu}) + \g^{\nu \lambda}  \g_{\beta \sigma} \partial_{\lambda} ( \delta  \g^{\sigma \beta} ) 
 - \g^{\mu \alpha} \partial_{\lambda} \g_{\mu \alpha} ( \delta \g^{\nu \lambda} ) .  
 \label{eq:AppBterm1}
 \end{equation}

Next consider the second term in the integrand of the l.h.s of eq.~\ref{eq:defH2}. Since  $ 2\Chris{\alpha}{\alpha}{\mu} = \g^{\alpha \lambda} \partial_{\mu} \g_{\alpha \lambda}$, 
 \begin{eqnarray}
 2 \delta \Chris{\alpha}{\alpha}{\mu} &=& (\delta \g^{\alpha \lambda}) \partial_{\mu} \g_{\alpha \lambda} 
 + \g^{\alpha \lambda} \partial_{\mu} (\delta \g_{\alpha \lambda}) \nonumber \\
&=& -(\delta \g^{\alpha \lambda}) \partial_{\mu} \g_{\alpha \lambda} -  \g_{\alpha \lambda}  \partial_{\mu} ( \delta \g^{\alpha \lambda} ) 
\label{eq:AppBterm2}
\end{eqnarray}

Using equations \ref{eq:AppBterm1} and \ref{eq:AppBterm2}, we find 
\begin{eqnarray}
 \B_{\nu}  \left[ \g^{\mu \alpha} \delta \Chris{\nu}{\mu}{\alpha} -  \g^{\mu \nu} \delta \Chris{\alpha}{\alpha}{\mu} \right]
\sqrt{-\mathbf{g}} 
&=& 
\B_{\nu} \left[- \partial_{\mu} (\delta \g^{\mu \nu} )  
  +  \g^{\mu \nu}  \g_{\alpha \beta} \partial_{\mu} ( \delta  \g^{\alpha \beta} )  \right]  \sqrt{-\mathbf{g}} 
 \nonumber \\
&+& \B_{\nu} \left[ - \g^{\nu \lambda} \partial_{\mu} \g_{\lambda \sigma} ( \delta \g^{\sigma \mu}  )
 + \frac{3}{2} \g^{\nu \lambda} \partial_{\lambda} \g_{\mu \sigma} (\delta \g^{\sigma \mu} ) 
   -\frac{1}{2} \g^{\mu \sigma} \partial_{\lambda} \g_{\mu \sigma} ( \delta \g^{\nu \lambda} )     \right]      \sqrt{-\mathbf{g}} 
 \\ \nonumber   
\end{eqnarray}
The first line in the r.h.s of the above equation contains terms with the derivatives of the variation. We note that these terms are eventually going to be integrated. By integrating them by parts and ignoring the boundary terms,  the above equation takes the form
\begin{eqnarray}
 \B_{\nu}  \left[ \g^{\mu \alpha} \delta \Chris{\nu}{\mu}{\alpha} -  \g^{\mu \nu} \delta \Chris{\alpha}{\alpha}{\mu} \right]
\sqrt{-\mathbf{g}} 
&=& 
 \left[ \partial_{\mu} (\B_{\nu}  \sqrt{-\mathbf{g}} ) \delta \g^{\mu \nu} 
  -  \partial_{\mu} (\g^{\mu \nu}  \g_{\alpha \beta} \B_{\nu}  \sqrt{-\mathbf{g}} )  \delta  \g^{\alpha \beta}   \right] 
 \nonumber \\
&+& \B_{\nu} \left[ - \g^{\nu \lambda} \partial_{\mu} \g_{\lambda \sigma} ( \delta \g^{\sigma \mu}  )
 + \frac{3}{2} \g^{\nu \lambda} \partial_{\lambda} \g_{\mu \sigma} (\delta \g^{\sigma \mu} ) 
   - \frac{1}{2} \g^{\mu \sigma} \partial_{\lambda} \g_{\mu \sigma} ( \delta \g^{\nu \lambda} )     \right]      \sqrt{-\mathbf{g}} 
\label{eq:AppBfin}
\end{eqnarray}

To simplify the r.h.s of eq.~\ref{eq:AppBfin} it is useful to note the following identities
\begin{equation}
\partial_{\mu}(\g^{\mu \nu}) = \g^{\sigma \nu} \g_{\sigma \lambda} \partial_{\mu }(\g^{\mu \lambda}) 
=  - \g^{\sigma \nu} \g^{\mu \lambda} \partial_{\mu}(\g_{\sigma \lambda}) .
\end{equation}
\begin{equation}
\partial_{\mu}(\sqrt{-\mathbf{g}}) = \frac{ \partial_{\mu} \mathbf{g} }{2 \mathbf{g}} \sqrt{-\mathbf{g}}
= \frac{\sqrt{-\mathbf{g}}}{2} \mathbf{g}^{i j} \partial_{\mu}(\mathbf{g}_{i j}) 
\end{equation}
\begin{eqnarray}
\mathbf{g}^{i j} \partial_{\mu}(\mathbf{g}_{i j}) &=& \g^{\sigma \lambda} \partial_{\mu} ( \g_{\sigma \lambda})  
+ \g^{\sigma \lambda} \partial_{\mu}( \epsilon \A_{\sigma} \A_{\lambda} \Phi^{2})
+ 2 \mathbf{g}^{\sigma 5} \partial_{\mu}(\epsilon \A_{\sigma} \Phi^{2})
+ \mathbf{g}^{55} \partial_{\mu}(\epsilon \Phi^{2}) \nonumber \\
&=& \g^{\sigma \lambda} \partial_{\mu} ( \g_{\sigma \lambda}) +  2 \J_{\mu}
\end{eqnarray}

Using the above identities, the r.h.s of eq.~\ref{eq:AppBfin} becomes  
\begin{eqnarray}
 \B_{\nu}  \left[ \g^{\mu \alpha} \delta \Chris{\nu}{\mu}{\alpha} -  \g^{\mu \nu} \delta \Chris{\alpha}{\alpha}{\mu} \right]
\sqrt{-\mathbf{g}} 
&=& \left[  \partial_{\mu} (\B_{\nu}) \delta \g^{\mu \nu} 
-  \partial_{\mu} (\B_{\nu}) \g^{\mu \nu}  \g_{\alpha \beta}  \delta  \g^{\alpha \beta}   \right. \nonumber \\
&+& \B_{\nu} \g_{\alpha \beta} \g^{\sigma \nu} \g^{\mu \lambda} \partial_{\mu} (\g_{\sigma \lambda}) 
\delta \g^{\alpha \beta} 
- \frac{1}{2} \B_{\nu} \g^{\mu \nu}  \g_{\alpha \beta} \g^{\sigma \lambda} \partial_{\mu} (\g_{\sigma \lambda}) \delta \g^{\alpha \beta} \nonumber \\
&-&  \B_{\nu} \g^{\nu \sigma} \partial_{\mu} (\g_{\sigma \alpha}) \delta \g^{\alpha \mu}
 + \frac{1}{2} \B_{\nu} \g^{\nu \sigma} \partial_{\sigma} (\g_{\alpha \mu}) \delta \g^{\alpha \mu}  \nonumber \\
&+&  \left.  \B_{\nu} \J_{\mu} \delta \g^{\mu \nu} -  \B_{\nu}  \J_{\mu}   \g^{\mu \nu}  \g_{\alpha \beta} \delta  \g^{\alpha \beta}  \right] \sqrt{-\mathbf{g}} .
\end{eqnarray}
Rewriting the derivatives of the metric in terms of Levi-Civita connection, we find  
\begin{eqnarray}
 \B_{\nu}  \left[ \g^{\mu \alpha} \delta \Chris{\nu}{\mu}{\alpha} -  \g^{\mu \nu} \delta \Chris{\alpha}{\alpha}{\mu} \right]
\sqrt{-\mathbf{g}}
&=& \left[  \partial_{\mu} (\B_{\nu}) \delta \g^{\mu \nu} 
-  \partial_{\mu} (\B_{\nu}) \g^{\mu \nu}  \g_{\alpha \beta}  \delta  \g^{\alpha \beta} \right.
\nonumber \\
&-&  \B_{\nu}   \Chris{\nu}{\mu}{\sigma}  \delta \g^{\sigma \mu}
+  \B_{\nu}  \Chris{\nu}{\mu}{\sigma}  \g^{\sigma \mu} \g_{\alpha \beta}  \delta  \g^{\alpha \beta}
 \nonumber \\
&+& \left.  \B_{\nu} \J_{\mu} \delta \g^{\mu \nu} 
 -  \B_{\nu}  \J_{\mu}   \g^{\mu \nu}  \g_{\alpha \beta} \delta  \g^{\alpha \beta} 
\right] \sqrt{-\mathbf{g}} . \nonumber
\end{eqnarray}
\begin{equation}
 = \left[ \nabla_{ \mu} \B_{\nu} - (\nabla \cdot \B ) \g_{\mu \nu}  + \J_{ \mu} \B_{\nu } - (\J \cdot \B) \g_{\mu \nu} \right] 
 \delta \g^{\mu \nu}  \sqrt{-\mathbf{g}}
\end{equation}

Since the variation $\delta \g^{\mu \nu}$ is symmetric in the indices $\mu$ and $\nu$, only the symmetric part of the r.h.s of the above equation will contribute to the equations of motion. Hence $\h_{\mu \nu}$ will be given by eq.~\ref{eq:defH3}.
  
\end{widetext}

\end{document}